\documentclass[twocolumn, twocolappendix]{aastex63}

\received{XXX}
\revised{YYY}
\accepted{ZZZ}
\submitjournal{ApJ}

\shorttitle{Origins of UHE CRs beyond the GZK horizon}
\shortauthors{Owen et al.}
\usepackage{amsmath}
\usepackage{tikz} 
\graphicspath{{./}{figures/}}
\newcommand{\myarrow}{\tikz\draw[thin,black,-latex] (0,2.5ex) -- ++(0,-1.5ex) -- +(2.5ex,0);}

\begin{document}

\title{Ultra high-energy cosmic rays from beyond the Greisen-Zatsepin-Kuz'min horizon}

\correspondingauthor{Ellis R. Owen, Qin Han, Kinwah Wu}
\email{erowen@gapp.nthu.edu.tw (ERO), mg1826002@smail.nju.edu.cn (QH), kinwah.wu@ucl.ac.uk (KW)}

\author[0000-0003-1052-6439]{Ellis R. Owen}
\affiliation{Institute of Astronomy, National Tsing Hua University, Hsinchu, Taiwan (ROC)}
\affiliation{Centre for Informatics and Computation in Astronomy, National Tsing Hua University, Hsinchu, Taiwan (ROC)}

\author[0000-0002-2342-9956]{Qin Han}
\affiliation{School of Astronomy and Space Science, Nanjing University, 
  163 Xianlin Avenue, Nanjing, Jiangsu 210093, 
   People's Republic of China}
\affiliation{Mullard Space Science Laboratory, University College London, Holmbury St Mary, Surrey RH5 6NT, UK}

\author[0000-0002-7568-8765]{Kinwah Wu}
\affiliation{Mullard Space Science Laboratory, University College London, Holmbury St Mary, Surrey RH5 6NT, UK}
\affiliation{Research Center for Astronomy, Astrophysics and Astrophotonics, Macquarie University, Sydney, NSW 2019, Australia}

\author[0000-0002-3349-3089]{Y. X. Jane Yap}
\affiliation{Institute of Astronomy, National Tsing Hua University, Hsinchu, Taiwan (ROC)}

\author[0000-0002-8516-6469]{Pooja Surajbali}
\affiliation{Max-Planck-Institut f\"{u}r Kernphysik, Saupfercheckweg 1, Heidelberg 69117, Germany}

\nocollaboration{5}



\begin{abstract}
Ultra-high-energy (UHE) cosmic rays (CRs) 
  of energies $\sim (10^{18}-10^{20})~{\rm eV}$, 
  accelerated in violent astrophysical environments,   
  interact with cosmic background radiation fields
  via photo-hadronic processes,  
  leading to strong attenuation. Typically, the Universe would become `opaque' to UHE CRs after several tens of Mpc, setting the boundary of the Greisen-Zatsepin-Kuz'min (GZK) horizon. 
  In this work, we investigate the 
  contribution of sources beyond the conventional GZK horizon to the UHE CR flux observed on Earth,
  when photo-spallation 
  of the heavy nuclear CRs is taken into account. We demonstrate this contribution is substantial, despite the strong attenuation of UHE CRs. 
A significant consequence 
  is
  the emergence of an 
  isotropic background component in the observed flux of UHE CRs, 
  coexisting with the anisotropic foreground component 
  that are associated with nearby sources. 
Multi-particle CR horizons, which evolve over redshift, 
  are determined by the CR nuclear composition. 
Thus, they are dependent on the source populations  
  and source evolutionary histories. 
\end{abstract}

\keywords{cosmic ray sources --- 
ultra-high energy cosmic radiation --- 
secondary cosmic ray --- 
astronomical radiation source --- 
extragalactic astronomy}


\section{Introduction} 
\label{sec:intro} 

Ultra high-energy (UHE)\footnote{ 
  \cite{Kachelries2019PrPNP} 
  referred to cosmic rays 
  with energies above $10^{17}\;\!{\rm eV}$ 
  as UHE cosmic rays. 
  The same terminology is adopted in this work.} 
  cosmic-ray (CR) events are rare.  
Although the detection of UHE CRs 
  with energies $E > 10^{20}\;\!{\rm eV}$   
  has been reported 
  \citep[e.g.][]{Bird1994ApJ, Bird1995ApJ}, 
  their flux $J(E)$ at sea level, 
  given by 
  $E^3 J(E) \approx 10^{24}\ 
   {\rm {eV^2}\;\! m^{-2}\;\! s^{-1}\;\! sr^{-1}}$,  
  at energy $E = 10^{20}\;\!{\rm eV}$ 
  \citep[see e.g.][]{Abu-Zayyad2013ApJ,Ivanov2017wH}, 
  implies an event rate of 
  only 1 particle per square km per century. 
Several origins of these UHE CRs have been proposed, 
  based on the argument that charged particles can,  
  in principle, be accelerated 
  to energies as high as $10^{20}\;\!{\rm eV}$ 
  in astrophysical environments, including 
  the relativistic jets of active galactic nuclei (AGN), 
  the large-scale shocks associated with galaxy clusters,
  and neutron stars,  
  as indicated by the~\cite{Hillas1984ARAA} criteria. 
  
High energy CR protons and nuclei interact with photons,  
 and these interactions lead  
 to the generation of lower energy subatomic particles.  
Cosmic Microwave Background (CMB) radiation and Extra-galactic Background Light (EBL)
  permeate the Universe, and these
  provide a target
  for UHE CR interactions, via photo-hadronic processes.  
Such photo-hadronic interactions attenuate the flux of CRs at high energies, 
   setting an energy limit above which 
   the CR flux detected at Earth would drop very substantially. 
This energy is around $\sim 10^{20}\;\!{\rm eV}$ 
  \citep{Greisen1966PhRvL,Zatsepin1966JETPL} and is known as 
   the Greisen-Zatsepin-Kuz'min (GZK) limit  
   \citep[For reviews of GZK limit, 
   see e.g.][]{Watson2014RPPh,Kachelries2019PrPNP}. 
Photo-hadronic interactions also limit the intergalactic distance 
  over which CR protons and nuclei can travel.  
For instance, the expected survival distance of a CR proton of 
  $E_{\rm p}\approx 3\times 10^{20}\;\!{\rm eV}$ 
  traversing through a blackbody radiation field 
  of a temperature $2.7\;\!{\rm K}$), 
  is $<30\;\!{\rm Mpc}$ 
  \citep[see e.g.][]{Kachelries2019PrPNP}.   
This has led to a belief that the extra-galactic UHE CR flux 
  with $E\sim 10^{20}~{\rm eV}$
  detected on Earth is likely comprised of CRs originating in nearby foreground sources within a horizon 
  (hereafter, referred to as the `GZK horizon') 
  of radius $\sim 50\;\!{\rm Mpc}$, and a residual, heavily-attenuated background flux from more distant source populations.\footnote{Contributions from source populations beyond $\sim$50 Mpc were considered in~\cite{Berezinsky1988AAP}, where UHE CR flux from sources up to $z\sim1.17$ were shown to provide a non-negligible component of the detected flux; see also~\cite{Kalashev2008JCAP} for more recent work considering the contribution from sources up to a few hundred Mpc.}  
 
If we accept the~\citet{Hillas1984ARAA} criteria strictly,
  the number of possible UHE CR sources 
  within the GZK horizon is limited.
Because of this and the close proximity of sources, 
   if the propagation of
   UHE protons and nuclei is ballistic\footnote{The propagation of charged UHE CRs
   is influenced by the strength and structure 
   of the magnetic fields in intergalactic space.   
   CRs of higher energy $E$ and/or lower charge $Z$ 
   (i.e. with higher rigidity $R = E/Z$) are deflected less. 
However, CRs with energies $>2\times 10^{19}~{\rm eV}$ 
  would experience little deflection ($\leq 10$ degrees) 
  over $\sim 100~{\rm Mpc}$ \citep{Batista2019FrASS}.}, 
  their arrival directions 
  as observed at Earth would not be isotropic.
Instead, they would be preferentially oriented 
   towards nearby UHE CR sources that fall within the GZK horizon.
Recent studies  
    \citep[for reviews, see][]{Kotera2011ARAA,Deligny2017PTEP,Batista2019FrASS, Rieger2019hepr} 
    have searched for anisotropies in the arrival directions of UHE CRs 
    \citep{Abreu2010APh, Aab2014ApJ, Aab2015ApJ, Aab2017Sci, Aab2018ApJ, Abbasi2018ApJ}, 
    with \cite{Aab2017Sci} 
    reporting a detection
    for CRs of energies 
    above $8\times 10^{18}\;\!{\rm eV}$ 
     at a significance $>5.2 \sigma$   
    (see section~\ref{sec:CR_anisotropy} for further discussion). 
Complementary studies investigating correlations 
   between UHE CR arrival directions and catalogues 
   of possible sources (in particular, AGN) 
   have also returned encouraging results  
   \citep[see][]{Abraham2007Sci, Abreu2010APh}. 
 However, stronger correlations result 
   if the UHE CR flux arriving at Earth is considered to have 
   two components: an anisotropic \textit{source} component 
   (presumably from nearby high-energy cosmic accelerators), 
   and an isotropic \textit{background} component~\citep{Kim2011JCAP}.
   
In this work, 
    we investigate the interactions of UHE CRs with the CMB and EBL,  
    considering in particular  
    how photo-spallation and photo-pion interactions affect the propagation and attenuation of UHE CR nuclei.
    We investigate how much of the UHE CR flux originates from distant sources, and assess its dependence on CR composition and source population models. We show that distant CR sources, together with secondary CRs produced by photo-spallation could
       supply a significant isotropic flux in the UHE CRs observed at Earth. 
The paper is organized as follows: 
  in {Section}~\ref{sec:propagation_interactions}, 
    we introduce our models for the propagation and interaction 
    of energetic CR nuclei 
    through intergalactic radiation fields, 
    accounting for the relevant micro-physics and particle interactions. 
In {Section}~\ref{sec:source_model}, 
  we introduce our CR source population, composition and spectral model. 
We present our results in {Section}~\ref{sec:results}    
   and summarize our findings and present conclusions 
   in {Section}~\ref{sec:conclusions}.

\section{Cosmological interactions and propagation of cosmic rays} 
\label{sec:propagation_interactions}

\subsection{Ultra high-energy cosmic ray interactions} 
\label{subsec:UHECR_interactions}

\subsubsection{Protons and neutrons}
\label{subsubsec:proton_channels}

Energetic protons interact with ambient radiation fields 
  to produce leptons and pions.  
These photo-hadronic interactions are dominated by 
  Bethe-Heitler (BH) photo-pair production 
  \citep{Bethe1934PRSA}  
  and photo-pion production processes. 
Photo-pair processes proceed as 
\begin{equation}%
\label{eq:BH1}%
{\rm A}' + \gamma \rightarrow {\rm A} + {l}^+ + {l}^- \ ,%
\end{equation}%
   where ${\rm A}'$ and ${\rm A}$ are nucleons  
   before and after the interaction respectively, 
   while ${l}^+$ and ${l}^-$ are the leptons and anti-leptons 
   produced in this process. 
As photo-electron pair production contributes most  
  to photo-pair CR energy losses 
   \citep[see e.g.][]{Blumenthal1970PRD,Klein2006RaPC} 
  over the range of energies of interest in this work\footnote{See \citet{Stepney1983MNRAS} for an analytical fit for the effective cross-section of this process, which we adopt in our calculations.}, 
  we consider only 
  the production of electrons (${e}^-$) and positrons (${e}^+$). 
Without loss of generality, 
  hereafter we do not distinguish between electrons and positrons 
  unless otherwise stated, 
  and both are referred to as ``electrons''. 
  
Photo-pion production 
  operates mostly through the following channels:
 (i) resonant single-pion production, 
 (ii) direct single-pion production, 
 and (iii) multiple-pion production 
   \citep{Mucke1999PASA}\footnote{Processes such as diffractive scattering (among others) may also operate, but are less important and do not arise at a rate comparable to 
   those of the dominant channels.}.  
Resonant single-pion production arises 
   through the production of $\Delta^+$ intermediaries,    
   which decay predominantly through the channels:  
\begin{align}%
\label{eq:pg_int}%
\rm{p} + \gamma &  \rightarrow \Delta^{+}  \rightarrow%
	\begin{cases}%
	    \rm{p} + \pi^0 \rightarrow \rm{p} + 2\gamma				\\[0.5ex]%
		\rm{n} + \pi^+ \rightarrow \rm{n} + \mu^+ + \nu_{\mu}		\\%
		\hspace{6.6em} \myarrow ~\rm{e}^+ +  \nu_{\rm e} + \bar{\nu}_{\mu}%
	\end{cases}  %
\end{align}%
  \citep[see][]{Berezinsky1993PRD}.  
The decays produce charged and neutral pions, 
  and their branching ratios (BRs) are roughly 2/3 and 1/3 respectively. 
The pions will further decay, 
  producing cascades of lower-energy particles.  
Multi-pion production occurs at higher energies 
  \citep[see][]{Mucke1999PASA}, 
  and when the invariant interaction energy exceeds $\epsilon_{\rm r} \approx 3500$,  
 charged and neutral pion production can arise with high multiplicities. 
For the energy range of interest in this work, 
   each type of pion is produced in roughly equal proportion when  
 all processes and the efficiencies of their production channels 
   are taken into account~\citep[see][]{Dermer2009book}. 
  
Calculations for interaction lengths 
  and their corresponding path lengths 
  in blackbody intergalactic radiation fields  
  were presented in \citet{Owen2018MNRAS} and \citet{Owen2019AA}.  
The same treatment is also adopted in this work. We note that free neutrons are produced 
  in photo-hadronic interactions, 
  as the one described in Equation~(\ref{eq:pg_int}), 
  and these neutrons will undergo a $\beta^{-}$-decay: 
\begin{equation}%
\rm{n} \rightarrow \rm{p} +  \rm{e}^- + \bar{\nu}_{\rm e} \ ,   %
\end{equation}%
   with a mean life of $880\;\!{\rm s}$ 
   \citep{Nakamura2010RPP}.   
A neutron will decay into a proton, an electron and an electron anti-neutrino 
  if it does not collide and interact with other particles.   
In astrophysical environments,  
  before $\beta^{-}$-decay occurs, 
  a neutron could interact with the radiation field,  
  leading to pion production: 
\begin{align}%
\rm{n} + \gamma & \rightarrow \Delta^0 \rightarrow %
		\begin{cases} %
			\rm{p} + \pi^-  \\[0.5ex]%
			\rm{n} + \pi^0%
		\end{cases} \ .   %
\end{align}%
The BRs for $\Delta^{0}\rightarrow\pi^-$ 
   and $\Delta^{0}\rightarrow\pi^0$ are 1/3 and 2/3 respectively  
   near the energy threshold of the interaction 
   \citep{Hummer2010ApJ}. 

The rates of photo-pion interactions for neutrons and protons  
   do not differ much. 
Thus, the interaction cross-section 
  and path length for high-energy neutrons 
  can be approximated 
  by the corresponding values for protons 
  \citep{Hummer2010ApJ, Romero2020Univ}\footnote{The total interaction cross-sections for protons and neutrons only begin to differ substantially at much lower energies than of interest here, below 140 MeV 
  \citep{Morejon2019JCAP}. 
At low energies, Thomson scattering and pair-production 
   is possible for protons 
   but is much weaker for neutrons~\citep{Gould1993ApJ}.}.    
Despite the short mean life for neutron decay, 
  neutrons can still traverse a long distance to collide with another particle or a photon  
  because of time dilation resulting from their relativistic motion. However, in this work, we seek to assess the propagation distances and spectral form of UHE CRs based on their nucleon number $A$, with their charge only being consequential to specify the injection spectrum of each species. We assume that the injection of $A=1$ UHE CRs is dominated by protons, while CR neutrons only emerge as the interaction products of primary CRs. As such, we model the interactions and propagation of UHE CR neutrons and protons in the same way and do not distinguish between the two species\footnote{This would only differ in the presence of large-scale intergalactic magnetic fields which would deflect and modify the propagation of relativistic protons, but not neutrons. Such effects are not expected to substantially affect our results, and these details are left to dedicated future studies.}. 

There are several consequences of the photo-pion processes. 
An isotropic cosmogenic neutrino component 
  has been attributed to decaying pions arising from 
  photo-pion interactions of CRs with intergalactic radiation fields 
  \citep{Allard2006JCAP,Rodrigues2021PhRvL}. 
Moreover, electrons produced in a photo-pion cascade,  
  together with those electrons formed by BH pair-production,  
  could up-scatter the ambient low-energy photons to $\gamma$-rays, 
  producing a diffuse $\gamma$-ray flux.  
In comparison with the $\gamma$-rays produced directly 
  from the decay of neutral pions (a product of the photo-pion process), 
  this Compton-scattered component is expected to be substantially weaker\footnote{Note that Compton scattered $\gamma$-ray \textit{cascades} can be substantial when initiated by high-energy $\gamma$-rays. In these cascades, energetic $\gamma$-ray photons initiate pair-production in low-energy cosmological radiation fields as they propagate, with the electron pairs Compton up-scattering ambient photons back to $\gamma$-ray energies, leading to a reprocessing effect on high-energy $\gamma$ radiation~\citep[e.g.][]{Dominguez2011MNRAS, Gilmore2012MNRAS, Inoue2013ApJ_atten, Owen2021MNRAS}. This is a separate process from the production of diffuse $\gamma$-rays by electrons formed in the photo-pion cascade, which are referred to instead as \textit{cosmogenic}.}.

\subsubsection{Nuclei}
\label{subsubsec:nuclei}

CR nuclei interact with radiation fields 
  via BH pair production, 
  (nuclear) photo-pion production
  and nuclear photo-spallation. BH pair production proceeds as 
\begin{align} 
\label{eq:BH2}%
{{\rm N}_1}' + \gamma & \rightarrow {\rm N_1} + {l}^+ + {l}^- \   
\end{align}  
\citep{Blumenthal1970PRD},     
  where ${{\rm N}_1}'$ and ${\rm N}_1$ are the nuclei 
  before and after the interaction respectively.   
The BH process is $Z^2$ times faster  
  for nuclei with charge number $Z$ than for protons.    
Photo-pion production channels are the same for nuclei as they are for protons
  (see Section \ref{subsubsec:proton_channels}), 
  with corresponding interaction products: for example, nuclear photo-pion processes 
   can also contribute to cosmogenic neutrino production.
However, during their interactions,
higher kinetic energies are available for nuclei compared to that for  protons.
The contribution of protons is typically weighted 
  by a factor of $A = m_{{\rm N}_1}/m_{\rm p}$
  \citep[e.g., see][]{Stecker1979ApJ,Dermer2009book},     
  which implies that heavier nuclei 
  would lose more energy during photo-pion interactions than lighter nuclei. 

The main nuclear photo-spallation channels are
\begin{align}
{\rm N}_1 \rightarrow
\begin{cases}  
\hspace*{0.1cm}    {\rm N}_2 + \rm{p}   \\ 
\hspace*{0.1cm}    {\rm N}_3 + \rm{2p}       
\end{cases} \ , 
\label{eq:photo_disintegration_x} 
\end{align}%
  where ${\rm N}_1$ is the primary nucleus  
  of an atomic mass $A$, 
  and ${\rm N}_2$ and ${\rm N}_3$ are the secondary nuclei 
  of atomic masses $A-1$ and $A-2$ respectively 
  \citep{Stecker1969,Puget1976ApJ,Stecker1999ApJ,Ahlers2011PhRvD}. 
Note that neutrons can also be emitted instead of protons in this process.
Other branches, which remove more than 1 or 2 nucleons, 
  may occur in a single interaction.  
These processes, however, occur with much lower probability 
  and are not included in our calculations. 
  
Photo-spallation (disintegration) 
  is dominated by the formation of the Giant Dipole Resonance (GDR) intermediary.
The process may be considered as a two-step mechanism \citep{Puget1976ApJ}, 
where the first stage involves the photo-absorption by a nucleus and the subsequent stage is the emission 
  of a single or multiple nucleons 
  from the nucleus via a (statistical) decay   
  \citep{Levinger1960}. 
Supposing that each of these emitted nuclei
  retain a fraction of $1/A$ of the primary nucleus 
  in a photo-spallation event,   
an approximate relation can be obtained from 
  the energy partition between nucleons as
\begin{align}
    \epsilon_{A-1} 
    & \approx \left(\frac{A-1}{A} \right)\;\! \epsilon_{A} \ , 
\label{eq:conserve}
\end{align}    
  such that the energy of the produced secondary nuclei can be estimated in an interaction 
  \citep[see][]{Dermer2009book}.  
Here, $\epsilon_{\rm A}$ and $\epsilon_{\rm A-1}$ 
    are, respectively, the dimensionless energies of the nuclei 
    before and after a photo-spallation event,
    with $\epsilon_{\rm A} =E_{\rm A}/m_{\rm e}c^2 
    = \gamma_{\rm A}(m_{\rm A}/m_{\rm e})$ and  
 a similar expression for $\epsilon_{\rm A-1}$. 
A consequence of the photo-spallation of heavy nuclei 
   is the production of lighter nuclei in large numbers.  
Each of these secondary nuclei 
  interacts with ambient radiation, 
  hence initiating a new chain of photo-spallation.   
 UHE CR nucleons thus litter traces of particles    
  along their paths when they propagate across the Universe.

\subsection{Cosmological propagation of energetic particles}
\label{sec:cosmological_prop}

The propagation of energetic particles 
  in intergalactic space 
  is described by the transport equation. 
  In the regime appropriate for this study, 
 it may take the following form:   
\begin{equation}
\label{eq:boltzmann}
\frac{\partial \mathcal{N}_A}{\partial\;\! t} = -  \;\!\nabla \cdot \mathcal{S}_A + \frac{\partial}{\partial\;\! \epsilon_A}
 \big[\;\! b \;\! \mathcal{N}_A \;\!\big] 
 + \mathcal{Q}_A - \Lambda_A \mathcal{N}_A  \ .
\end{equation} 
  Here, $\epsilon_A$ is the energy (of the particles) and  
  $b$ is the rate of energy loss 
  (presumably caused by cooling).  
The index $A$ denotes the nuclear species. In this work, 
  we relate nuclear charge to mass following the correspondence presented in~\cite{Puget1976ApJ}. 
  The charge of a species is not strictly set unambiguously by its mass. However, in most cases, there is a clear relation between the mass number of an UHE CR and its species/charge, with alternative isotopes being unstable and not being preferentially formed\footnote{This is with the exception of $A=5$ and $A=8$, which do not form stable isotopes. Since the radioactive decay timescale is less than the one-nucleon photo-spallation timescale for all but $^{53}$Mn, $^{26}$Al and $^{10}$Be, we assume that the decay brings the secondary nucleus to the line of nuclear stability before the next photo-collision~\citep{Stecker1999ApJ}. As such, we consider the abundances of nuclei with $A=5$ and $A=8$ to be zero, and that decays which would yield such products would instead undergo double nucleon photo-spallation to the next stable species in the decay chain.}.  We further define $\mathcal{N}_A$ as the number density of the particles with mass number $A$, while   
$\mathcal{S}_A$ is the CR streaming term, 
  which is given by $c\mathcal{N}_A$
  in the observer's frame 
  if diffusion is negligible~\cite[e.g.][]{Webb1979ApSS}. 
$\mathcal{Q}_A$ is the source term  
  specifying the rate of particle injection.  
$\Lambda_A \mathcal{N}_A$ is the sink term 
  specifying the rate of particle loss, 
  with an efficiency set by the parameter $\Lambda_A$.  
In cosmological settings, 
  we assign $\mathcal{N}_A$
  as the co-moving number density 
  of a particle species.  
Therefore, $\mathcal{N}_A = n_A (1+z)^{-3}$, 
  where $n_A$ is the physical number density of a particle species,
  and $z$ is the cosmological redshift. 

The streaming of UHE CRs is effectively 
  the speed of light, $c$. 
Thus, Equation (\ref{eq:boltzmann}) becomes   
\begin{equation}
\label{eq:reduced_boltzmann}
\frac{\partial\;\! n_A}{\partial\;\! t} = -c \frac{\partial\;\! n_A}{\partial\;\! s} + \frac{\partial}{\partial\;\! \epsilon_A} 
 \big[\;\!b\;\! n_A  \;\!\big] 
   + (1+z)^3 \mathcal{Q}_A - \Lambda_A\;\! n_A  \ ,     
\end{equation} 
  with $s$ as the comoving distance. 
In terms of the cosmological redshift, this may be expressed as  
\begin{align}
\label{eq:transport_equation_general}
\frac{\partial}{\partial\;\! t}
\; n_A(\epsilon_A, z)  
 &= - c\;\!\left(\frac{{\rm d} s}{{\rm d} z}\right)^{-1}\;\! \frac{\partial}{\partial z}\;\! n_A(\epsilon_A, z) \nonumber \\  
 & \hspace*{0.6cm} + 
 \frac{\partial}{\partial\;\! \epsilon_A} 
  \big[\;\!b_A(\epsilon_A, z)\;\! n_A(\epsilon_A, z)\;\!\big]  
   \nonumber   \\ 
 &\hspace*{0.6cm} + {Q}_A(\epsilon_A, z) 
 - \Lambda_A\;\! n_A (\epsilon_A, z) \ ,
\end{align}
  where $Q_A = \mathcal{Q}_A (1+z)^{3}$. 
In a Friedmann-Lema{\^i}tre-Robertson-Walker (FLRW) universe, 
\begin{equation}
 \frac{{\rm d}s}{{\rm d}z} 
   = \frac{c~\mathcal{E}(z)}{H_0\;\!(1+z)} \ ,  
\end{equation}
  and  
\begin{equation}
\label{eq:e_func}
    \mathcal{E}(z) = \left[\Omega_{\rm r, 0}(1+z)^4 + \Omega_{\rm m, 0}(1+z)^3 + \Omega_{\rm \Lambda, 0} \right]^{-1/2}
\end{equation}
\citep[see, e.g.][]{Peacock1999_book},  
  where $\Omega_{\rm m,0} = 0.315\pm0.007$, 
  $\Omega_{\rm r,0} \approx 0$ 
  and $\Omega_{\rm \Lambda,0} = 0.685\pm0.007$ 
  are the normalised density parameters for matter, 
  radiation and dark energy respectively.  
The present value of the Hubble parameter 
  $H_0 = 100\;\! h\;\! {\rm km}\;\!{\rm s}^{-1}{\rm Mpc}^{-1}$,
  where $h = 0.673\pm 0.006$ 
  \citep{Planck2018}.  

A quasi-steady condition is adopted 
  in our calculations, 
  and this greatly simplifies the procedures involved 
  in solving the particle-transport equations. 
The quasi-steady condition is justified,  
  as the time of propagation of the particles  
  across length-scales of the order of their horizons 
  is much shorter than the Hubble time.  
Setting $\partial_t = 0$,  
  Equation~(\ref{eq:transport_equation_general}) becomes   
\begin{align}
\label{eq:transport_equation_steadystate}
\frac{\partial\;\! {n_A}}{\partial z} 
  =  \frac{1}{c}
    \left[\frac{\partial}{\partial \epsilon_A} 
    \left(b_A n_A \right)+ {Q}_A -  \Lambda_A n_A \right]
    \frac{{\rm d} s}{{\rm d} z} \ ,  
\end{align}
  which will be used    
  in {Section}~\ref{subsec:protons_nuclei}  
  for the transport calculations of CR protons and nuclei.

\subsubsection{Interaction rates}
\label{subsec:protons_nuclei}

The source term  in Equation (\ref{eq:transport_equation_steadystate})  
  has two components,
\begin{equation}
   Q_A = Q^{\rm a} + Q^{\rm sp} \   .
\end{equation} 
  Here, 
  $Q^{\rm a}$ specifies the rate primary nuclei are supplied by a source population of cosmic accelerators, and is defined by the source population, its evolution (specified by redshift, $z$),  
  and the CR energy $\epsilon_A$ and mass number $A$ 
  (our adopted models are discussed in {Section}~\ref{sec:source_model}). 
$Q^{\rm sp}$ is the rate of production 
  of secondary nuclei from photo-spallation 
  of heavy nuclei (i.e. the particle cascade contribution).
We consider that 
  the injected particles are represented by  $^1$H, $^4$He, $^{14}$N, $^{28}$Si and $^{56}$Fe  
  (with their relative abundance fractions detailed 
  in {Section}~\ref{sec:source_model}). 
The spallation (disintegration) of heavy CR nuclei 
  produces secondary nuclei with mass numbers 
  ranging from 55 to 1, and 
the injection rate of secondary nuclei (for mass number less than 56) 
  by photo-spallation particle cascades 
  is given by  
\begin{equation}
{Q^{\rm sp}}_{\! A}(z, \epsilon_A) 
   = \sum_{{A}'}
   \int_{\epsilon_{\rm min}}^{\epsilon_{\rm max}}
   {\rm d}\;\!\epsilon_{\rm {A}'}\; 
   \frac{{R}^{{A}'}_A \;\! n_{{A}'}}{(1+z)^3} \ .   
   \label{eq:full_sp_rate}
\end{equation} 
Here, $\epsilon_{\rm min}$ and $\epsilon_{\rm max}$ 
  are the minimum and maximum energies considered in our calculations.
  We set the lowest energy to be 
  $\epsilon_{\mathrm{min}} m_{\rm e} c^2 = 3.98 \times 10^{18}\;\!{\rm eV}$, as it is not clear 
  if the cosmic ray spectrum is dominated by extra-galactic particles 
  at lower energies than this~\citep{Giacinti2012JCAP, Aloisio2014JCAP}.
The maximum energy 
  is set to be 
   $\epsilon_{\mathrm{max}} m_{\rm e} c^2 = 3.16\times 10^{20}\;\!{\rm eV}$, corresponding to the most energetic UHE CRs 
   detected at Earth -- see~\citealt{Bird1995ApJ}. 
The summation in ${Q^{\rm sp}}_{\!A} (z, \epsilon_A)$  
  accounts for the primary nuclei from ${A}'$ up to $A=56$, 
  where the photo-spallation injection rate for the most massive species under consideration (i.e. ${Q^{\rm sp}}_{\!56}$) is zero. 
${Q^{\rm sp}}_{\!A}$ is dependent on $n_{{A}'}$ (where ${A}'>A$).  
As such, we solve 
  Equation~(\ref{eq:transport_equation_steadystate})   
  from the CR nuclear species with largest $A(=56)$  to those with the smallest $A$($=1$ for protons/neutrons).  
${R}^{{A}'}_A(\epsilon_{{A}'}, \epsilon_A, z)$ 
   is the differential rate 
   of the photo-spallation 
   of species with mass number 
   ${{A}'}$ into species with mass number $A$, 
   in the lab frame,  
   in the presence of a soft photon field 
   $n_{\rm ph}(\epsilon_{\rm ph})$.  
If the radiation field is considered to be isotropic, 
  this can be expressed as   
\begin{align}
\label{eq:reaction_rate}
R^{{A}'}_A &  = \frac{c}{2\gamma_A^2} 
   \int_0^{\infty} {\rm d}\epsilon_{\rm ph}\;\!  
   \Bigg\{ \;\! \frac{n_{\rm ph}(\epsilon_{\rm ph})}{ {\epsilon_{\rm ph}}^2}  
   \nonumber \\ 
 & \hspace*{1.75cm}\times  \int_0^{2 \gamma_{{A}'} \epsilon_{\rm ph}}
  {\rm d}\epsilon_{\rm r} 
  \;\!\epsilon_{\rm r} \sigma_{{{A}'}\rightarrow A}(\epsilon_{\rm r})\;\! \bigg\} 
\end{align}
\citep[see][]{Dermer2009book}. 
Here, $\gamma_{A'}$ is the Lorentz factor of the primary nucleus,  
  $\sigma_{{A'}\rightarrow A}(\epsilon_{\rm r})$ 
   is the photo-spallation  cross-section of the 
   primary nucleus of $A'$ to produce a secondary nucleus of $A$,  
    and $\epsilon_{\rm r}$ 
   is the total invariant interaction energy (in a dimensionless form).    
For a blackbody radiation field\footnote{The 
  photon number density of a blackbody radiation field 
  may be expressed as 
$n_{\rm ph}(\epsilon_{\rm ph}) 
  = f\;\!({8\pi}/{\lambda_{\rm c}}^3) 
    \{ {{\epsilon_{\rm ph}}^2}/
     [\exp (\epsilon_{\rm ph}/{\Theta})-1] \}$,   
  where $\lambda_{\rm c}$ is the Compton wavelength of an electron 
  and $\Theta = k_{\rm B} T/m_{\rm e}c^2$. $f=1$ for an unmodified black-body radiation field, e.g. the CMB, but $f<1$ in the case of a diluted radiation field (e.g. EBL components).},  
  the integrals 
  in Equation (\ref{eq:reaction_rate})   
   can be evaluated analytically. 
This simplifies the calculations 
  for the region 
  where the CMB dominates the radiation field, or the EBL if it is approximated as a summation of blackbody spectra (see Appendix~\ref{sec:appendixa}). 

In the two-step mechanism for photo-spallation, 
   proper accounting for the physical processes in both steps  
   are required 
   in order to obtain an accurate value for the cross-section 
   \citep{Puget1976ApJ}. 
For the case of single nucleon emission, 
  the cross-section can, however, be adequately approximated 
  by a Lorentzian:  
\begin{equation}
\label{eq:lorentzian_cross_section}
    \sigma_{1}(\epsilon_{\rm r})
    = \frac{\sigma_{\rm 0, A}{\epsilon_{\rm r}}^2{\Delta_{n}}^2}
    {({\epsilon_0}^2-{\epsilon_{\rm r}}^2)^2 + {\epsilon_{\rm r}}^2 {\Delta_{n}}^2}
\end{equation}
\citep[see][]{Karakula1993APh,Wang2008ApJ}. Photo-spallation may result in multiple nucleon emission
  when $\epsilon_{\rm r}$ is above 30~MeV 
  or below the pion-production threshold ($\simeq 150$~MeV), 
  albeit with a much smaller cross-section and lower rate. 
A more complicated expression is available  
 to describe this variation in the cross-section 
 \citep[see][]{Puget1976ApJ},  
 which is appropriate 
 when the ambient radiation field has a very hard spectrum 
 \citep[see][]{Wang2008ApJ}.  
However, for GZK {attenuation} calculations, 
  the ambient radiation {fields} which dominates the interactions {are the CMB and EBL, for which} the 
 simplified expression for the cross-section 
is sufficient.

For analytical tractability, this formulation can be simplified further 
  by replacing the Lorentzian with a delta-function:
\begin{equation}
    \label{eq:delta_cross_section}
    \sigma_{1}(\epsilon_{\rm r}) = \sigma_{\rm 0, A}\;\! \Delta_n \delta(\epsilon_{\rm r} - \epsilon_0) \   
\end{equation}     
  \citep{Wang2008ApJ, Dermer2009book}. This still yields  a photo-spallation rate correct to within an order-of-magnitude \citep{Wang2008ApJ}, which is sufficient for the purposes of our demonstrative model. 
In Equation (\ref{eq:delta_cross_section}), 
  the width of the cross-section in energy 
   $\Delta_{n} = 8 ~\text{MeV}$, 
     is given by the energy bandwidth of the giant resonance,  
    and the (dimensionless) energy at which the giant resonance cross-section peaks  
     is given by $\epsilon_0 = 83.46\;\! A^{-0.21}$ for $A>4$, 
    and $\epsilon_0 = 1.81 A^{2.433}$ for $A\leq 4$ 
 \citep{Karakula1993APh}. 
The maximum value of the cross-section is 
  $\sigma_{\rm 0, A} = 1.45 \times 10^{-27} A ~{\rm cm}^2$ 
  \citep{Wang2008ApJ}.  
Substitution of this cross-section, given in 
  Equation (\ref{eq:delta_cross_section}),    
   into Equation (\ref{eq:reaction_rate})  
   yields 
\begin{align}
\label{eq:approx_photo_R}
 R_1 & 
     = - \frac{4\pi \;\!f(z) \;\! c \sigma_{\rm 0,A} \Delta_n \epsilon_0 \Theta(z)}{{\gamma_A}^2{\lambda_c}^3} \nonumber \\ 
      &\hspace*{1cm} \times 
       \ln\left[\;\!1-\exp\left(\frac{-\epsilon_0}{2\gamma_A \Theta(z)}\right)\right] \  
\end{align}  
\citep[see also][for an expression derived under similar circumstances]{Dermer2009book}. 
Here, $\lambda_{\rm c}$ is the Compton wavelength of electron, $\Theta = k_{\rm B} T/m_{\rm e}c^2$ is
  the dimensionless temperature of the radiation field, and
  $f=1$ for a blackbody radiation field, e.g. the CMB, but $f(z)<1$ in the case of diluted blackbody radiation (e.g. redshift-dependent EBL components -- see Appendix~\ref{sec:appendixa}).
This rate can also be used 
  to compute the specific absorption rate of UHE CR nuclei 
  due to photo-spallation. 
For this, we adopt the single-nucleon emission approximation 
  of \cite{Wang2008ApJ}, giving  
\begin{align} 
{\Lambda^{\rm sp}}_{\!A} & \approx \frac{R_1}{A} \ .
\end{align}  
This forms a component  
  of the sink (or absorption) term 
  in the transport equation, i.e  
  Equation~(\ref{eq:transport_equation_steadystate}).
 
Adopting the single nucleon approximation 
  simplifies Equation ~(\ref{eq:full_sp_rate})  to  
\begin{align}
{Q^{\rm sp}}_{\! A}(z, \epsilon_A) &   \approx 
    \frac{1}{(1+z)^3} \int_{\epsilon_{\rm min}}^{\epsilon_{\rm max}} {\rm d}\epsilon_{{A}'} {R_1 n_{{A}'}}
\end{align}   
   for nuclei with mass number $A>1$, and  
\begin{align} 
\label{eq:single_nucleon_production} 
{Q^{\rm sp}}_{\! 1}(z, \epsilon_1)
  & \approx \frac{1}{(1+z)^3} \ 
  \Biggl\{2\int_{\epsilon_{\rm min}}^{\epsilon_{\rm max}} 
  {\rm d}\epsilon_{2} {R_{1} n_{2}}  \nonumber \\ 
  & \hspace*{2cm} +  \sum_{{A}'} \int_{\epsilon_{\rm min}}^{\epsilon_{\rm max}} 
  {\rm d}\epsilon_{{A}'} {R_1 n_{{A}'}} \Biggr\} \   
\end{align} 
  for $A=1$ nuclei.  
Here we do not distinguish between protons and neutrons 
  (see {Section}~\ref{subsubsec:proton_channels}), 
  and the first term in 
  Equation (\ref{eq:single_nucleon_production})  
  accounts for the photo-splitting of $A=2$ nuclei into two individual nuclei.

Nuclei can also be absorbed by photo-pion interactions.
The expression of the interaction reaction rate in this case 
    takes the same form 
    as that given by 
    Equation (\ref{eq:reaction_rate}), 
    but with the substitution 
    of the pion-production cross-section, 
    $\sigma^{A\pi}$. 
In this work, we also approximate this with a delta-function   
\begin{equation}
\label{eq:pion_cross_section}
  \sigma^{A\pi} 
   = \frac{ \pi \sigma_0  \Gamma_{\rm res} 
   A}{2}\ \delta(\epsilon -\epsilon_{\rm res}) \  ,
\end{equation}
 where $\Gamma_{\rm res}$ is introduced as the resonance width.
Adopting the same parameterisation 
  of $\epsilon_{\rm res}$ and $\sigma_0$  as~\citet{Unger2015PhRvD}\footnote{This was based on the computed cross-section values in~\citet{Kampert2013APh, Batista2013ICRC} using tools provided online at~\url{https://github.com/CRPropa/CRPropa3-data}.} 
\citep[see also, e.g.][]{Moncada2017JHEAp},   
  we obtain $\epsilon_{\rm res} \approx 665$, 
  $\sigma_0 = 5.0\times 10^{-28} {\rm cm^2}$,  and 
  $\Gamma_{\rm res}=150~{\rm  MeV}$.
By substituting Equation~(\ref{eq:pion_cross_section}) 
  into Equation~(\ref{eq:reaction_rate}) 
  we obtain an analytical expression 
  for the interaction rate:
\begin{align}
\label{eq:rate_pA}
{{R}^{A\pi}}_{\! A}
  &= \frac{4 \pi \;\!f\;\! c}{{\gamma_{A}}^2 {\lambda_c}^3} \int_0^{\infty} {\rm d}\epsilon_{\rm ph}\;\! 
  \bigg\{
  \frac{1}{ \exp\left[\epsilon_{\rm ph}/\Theta\right]-1} \nonumber \\
&\hspace{1.cm} \times \int_0^{2 \gamma_{A} \epsilon_{\rm ph}} {\rm d}\epsilon_{\rm r} \;\!\epsilon_{\rm r}\pi \sigma_0 \frac{ \Gamma_{\rm res}}{2}
  A\;\! \delta(\epsilon_{\rm r} -\epsilon_{\rm res})\bigg\} \nonumber \\ 
&=- \frac{2\pi^2 \;\!f\;\! c \sigma_0 \Gamma_{\rm res}A \epsilon_{\rm res} \Theta}{{\gamma_{A}}^2{\lambda_c}^3} \nonumber \\  
&\hspace*{2cm} \times \ln\left[1-\exp\left(\frac{-\epsilon_{\rm res}}{2\gamma_{A} \Theta}\right)\right] \ , 
\end{align} 
   where the last line 
   expresses the rate in a similar 
   form to that for photo-spallation, 
   which is given in Equation~(\ref{eq:approx_photo_R}). 
The corresponding specific absorption rate of CRs due to photo-pion interactions then follows simply as
\begin{equation} 
\label{eq:absorption_pion}
  {\Lambda^{A \pi}}_{\! A}=\frac{{{R}^{A\pi}}_{\! A}}{A} \ ,
\end{equation} 
such that the overall sink term for nuclei,  
 for use in Equation~(\ref{eq:transport_equation_steadystate}), 
 may be written as 
\begin{equation}
    \Lambda_A={\Lambda^{\rm sp}}_{\! A}+{\Lambda^{A \pi}}_{\! A} \ . 
\end{equation}  

The other CR interactions described in {Section}~\ref{subsec:UHECR_interactions} typically operate as continuous loss processes, or such that only a small fraction of the particle's energy is transferred in a single interaction. These can be modelled as cooling processes and the total cooling rate experienced by an UHE CR,   
    in terms of the dimensionless energy of the nucleon 
    (again, we retain our earlier convention where energies are in units of electron rest mass), therefore 
    consists of three components:
\begin{equation}
  b  = -{\rm d}\epsilon/{\rm d}t 
   = {b^{\gamma A}}_{\! A}+b_{\rm rad}+b_{\rm ad} \ ,   
\end{equation} 
where ${b^{\gamma A}}_{\! A}$ is the cooling rate 
  due to photo-pair production, 
  $b_{\rm rad}$ is the radiative cooling rate, 
  and $b_{\rm ad}$ is the adiabatic cooling rate of UHE CR nuclei as they propagate through an expanding cosmology.
 
In our calculations for photo-pair production, 
  the invariant dimensionless interaction energy $\epsilon_{\rm r} \gtrsim 60$, 
  so a fitted approximation for the inelastic cross-section of photo-pair losses 
  is available 
  \citep[see][]{Jost1950PR, Bethe1954PR, Blumenthal1970PRD, Stepney1983MNRAS}:  
\begin{equation}
\label{approx-sigma}
   {\sigma}^{\rm \gamma A}(\epsilon_{\rm r})\approx %
  \left\{	\frac{7}{6\pi}\;\! \alpha_{\rm f}
  \ln\left[\frac{\epsilon_{\rm r}}{k_{\rm \gamma e}}\right] \right\}
  {\sigma_{\rm T}} \ , 
\end{equation}%
where $\alpha_{\rm f}$ is the fine structure constant, 
  $\sigma_{\rm T}$ is the Thomson cross-section, and $k_{\rm \gamma e}$ 
  is a fitting constant which we set as 2.0~\citep[see][]{Owen2018MNRAS}. 
The CR cooling rate due to photo-pair losses can then be expressed as 
\begin{align}
\label{eq:photo_pion}
{b^{\gamma A}}_{\! A} & = - \frac{{\rm d}\;\!\epsilon_A}{{\rm d}\;\!t}
  \nonumber \\
  & = - \frac{f(z)\;\!Z_A^2\;\! c}{2\gamma_A A} \int_{{\gamma_A}^{-1}}^{\infty} {\rm d}\epsilon_{\rm ph}\;   
    \bigg\{
    \frac{ 1 }{\exp\left[\epsilon_{\rm ph}/\Theta(z)\right]-1} 
   \nonumber  \\ 
  &\hspace*{1.8cm} \times \int_2^{2 \gamma_A \epsilon_{\rm ph}} 
  \frac{{\rm d}\epsilon_{\rm r} \;\!\epsilon_{\rm r}{\sigma}^{\rm \gamma A}
  (\epsilon_{\rm r})}{\sqrt{1+2\epsilon_{\rm r}}} \bigg\} 
\end{align} 
\citep{Dermer2009book, Aloisio2013APh_CMBonly}, 
  where the lower boundary value of the inner integral follows from at least one pair of electrons being formed by the process.
As in \cite{Owen2019AA}, 
  we assume that the electron-positron pairs are produced 
  in the zero-momentum frame of the interaction, 
  and the interaction energy is completely 
  dominated by the contribution from the interacting nucleus    
  \citep[see also][]{Protheroe1996APh, Dermer2009book}. 
This yields  
\begin{equation}%
\label{eq:photopair_losses}%
 {b^{\rm \gamma A}}_{\! A}(\epsilon_A) 
\approx
	- \frac{f\;\!Z_A^2\;\!{c} \alpha_{\rm f} \sigma_{\rm T}\mathcal{F}_{\rm \gamma e}(u)}{{\lambda_{\rm c}}^3 {\gamma_{A}}^2 A\;\! u^{5/2}}   \ ,  
\end{equation}%
where the dimensionless variable $u =  (\gamma_A \Theta)^{-1}$, 
 and the function $\mathcal{F}_{\rm \gamma e}(u)$   
 takes the same form as that in \citet{Owen2019AA}. 
 
Radiative cooling is dominated by Compton scattering 
  with the ambient radiation field. It arises at a rate of 
\begin{align}%
\label{eq:radiative_losses}%
b_{\rm rad}(\epsilon_A, z) 
  &= - \frac{4}{3}\frac{\sigma_{\rm T}\;\! U(z)}{m_{\rm e} c} 
  \left[
   \frac{{\epsilon_A}^{1/2}Z_A}{A} 
   \left(\frac{m_{\rm e}}{m_{\rm p}}\right)      
   \right]^4 \  
\end{align}%
\citep[cf. the expression given in][]{Puget1976ApJ},   
where $U(z)$ is the energy density of the radiation field 
(at redshift $z$).
Note that, here, $b_{\rm rad}$ 
  is dependent on the charge of the nuclei $Z_A$.
  
Adiabatic losses due to cosmological expansion occur at a rate 
\begin{align}%
\label{eq:adiabatic_losses}%
  b_{\rm ad}(\epsilon_{A}, z) 
  &= - \epsilon_{A} \;\! H_{\rm 0}\;  \mathcal{E}(z) 
\end{align}%
\citep[see][]{Gould1975ApJ, Berezinsky1988AAP, Berezinsky2006PRD},
 where $\mathcal{E}(z)$ is given by 
 Equation (\ref{eq:e_func}).

\begin{figure}
    \centering
    \includegraphics[width=\columnwidth]{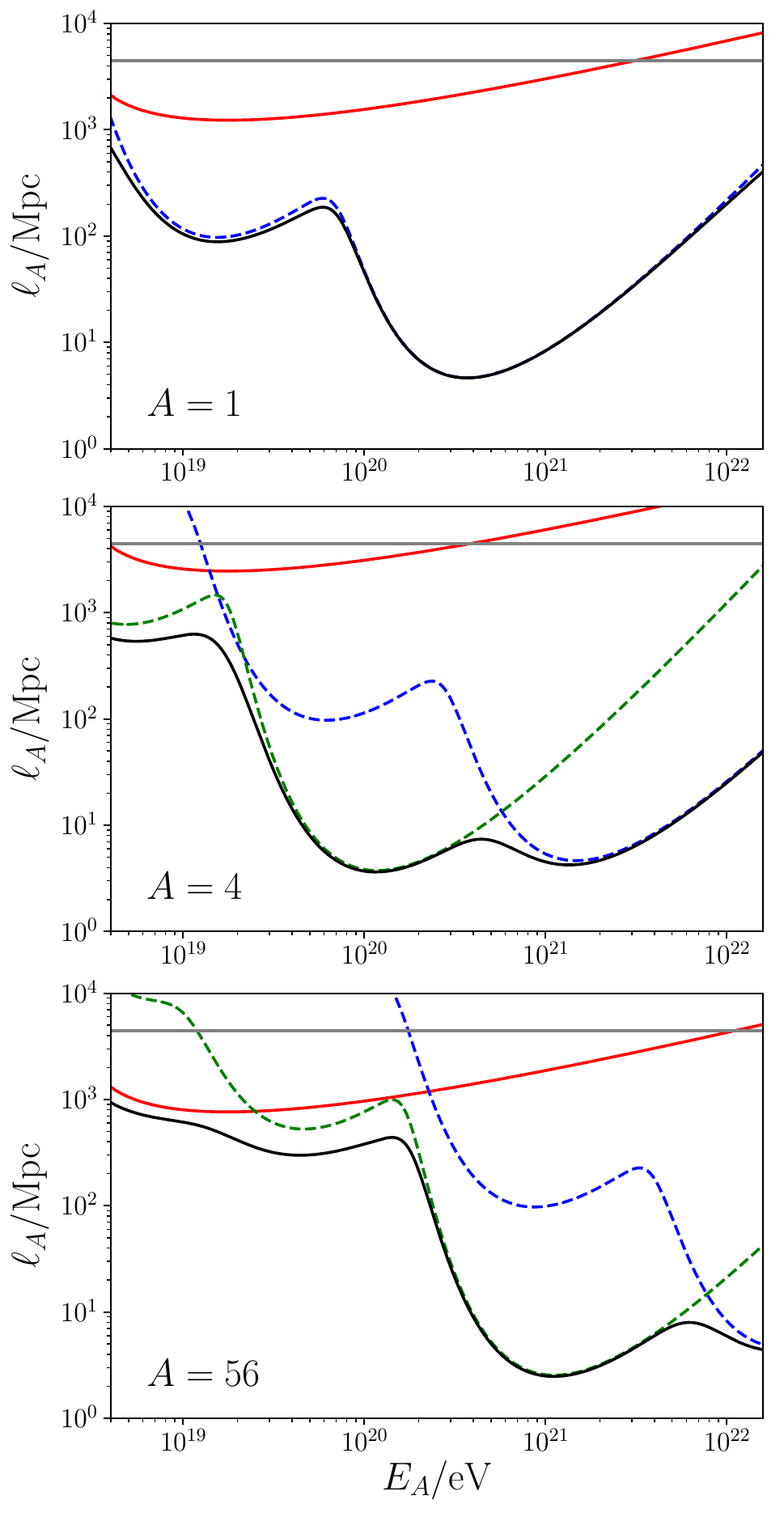}
\caption{The characteristic lengths of energy-loss   
  for nuclei traversing through background (CMB and EBL) cosmological radiation fields 
  at $z=0$, for various cooling and absorption processes. 
Panels from top to bottom correspond 
  to protons ($^1$H) , Helium ($^4$He) and Iron ($^{56}$Fe) CR nuclei. 
Adiabatic losses due to cosmological expansion 
  are represented by the grey lines; 
  pair production by the red curves; 
  photo-pion production by the blue dashed curves;   
  and photo-spallation by the green dashed curves. 
Here, solid lines and curves denote 
  continuous cooling loss, 
  and dashed curves denote 
  absorptive cooling loss. 
The solid black curves 
  represent the total cooling. }
  \label{fig:H_1_z0_all}
\end{figure}

\begin{figure*}
    \centering
    \includegraphics[width=0.9\textwidth]{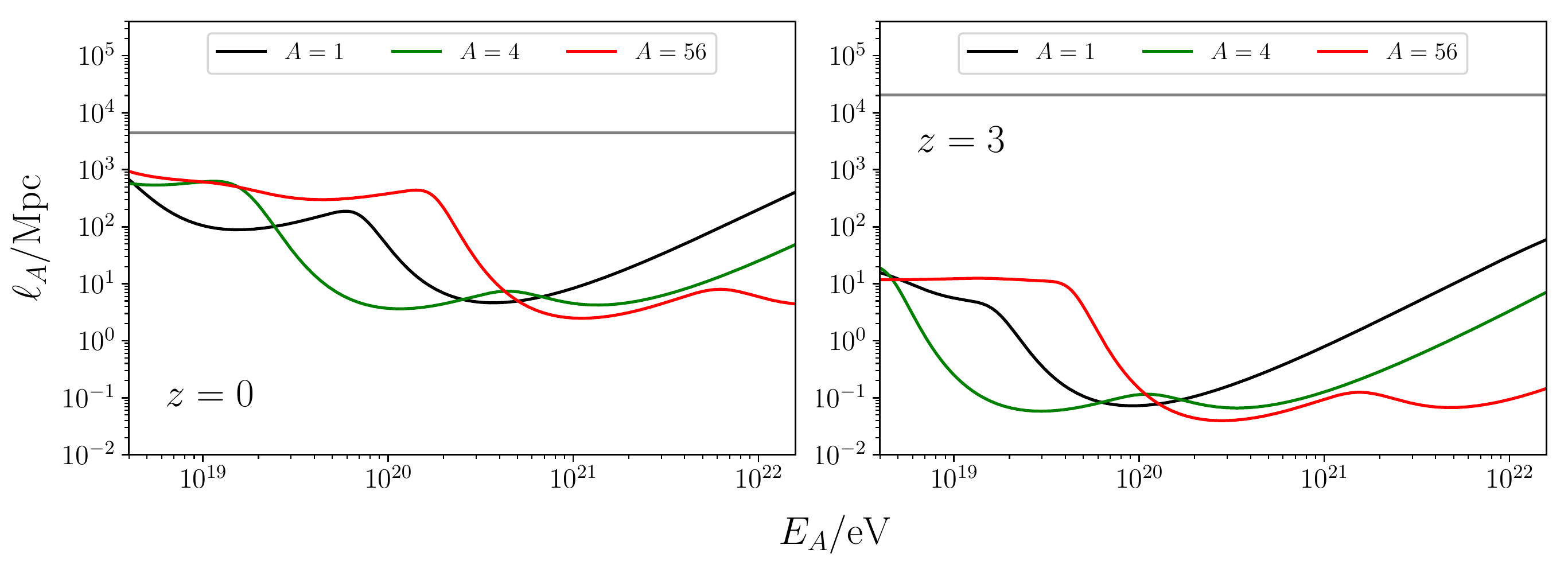}   
        \caption{ 
        Total characteristic energy-loss distances 
        of CR protons ($^1$H, black curves), 
        Helium ($^4$He, green curves) 
        and Iron ($^{56}$Fe, red curves) 
        in background (CMB and EBL) cosmological radiation fields 
        at $z=0$ (left panel) and $z=3$ (right panel).  
The corresponding energy-loss distances      
  for adiabatic cooling 
  due to cosmological expansion (grey lines) 
  are shown for comparison. 
The cooling losses are stronger at $z=3$, 
  because of a strong CMB radiation field, 
  but the adiabatic losses are less severe.}
  \label{fig:z3z0compare}
\end{figure*}

We compare the relative importance 
  of each of these cooling and interaction processes in {Figure}~\ref{fig:H_1_z0_all}, 
  where the effective path length
 of an UHE CR nucleus ($\ell_A$) is defined as the characteristic distance over which the CR loses its energy. 
 This can be estimated from the cooling rate $b$ using $\ell_A = c\;\![|b|/\epsilon_{A}]^{-1}$. 
In {Figure}~\ref{fig:H_1_z0_all}, 
   we show the effective loss lengths for UHE CR nuclei ($A=1$ for protons, $A=4$ for $^4$He nuclei and $A=56$ for $^{56}$Fe nuclei) 
   in intergalactic conditions at $z=0$, 
   with continuous losses due to pair production and adiabatic losses 
   shown by the solid lines. 
These are compared to conditions at $z=3$ 
  in {Figure}~\ref{fig:z3z0compare}, 
  where the background radiation fields are more intense 
  (leading to faster photo-pair losses and higher photo-pion and photo-spallation rates, but lower adiabatic loss rates due to the relatively lower rate of cosmic expansion at $z=3$ compared to the present Universe). 
Radiative losses are inconsequential in all cases.
For the purposes of comparison, 
  we also estimate corresponding effective path lengths for UHE CRs 
  due to processes which we regard as absorption in our calculations, 
  where a substantial amount of energy is lost by the CR in a single interaction.  
These are photo-pion production and photo-spallation 
  -- note that photo-spallation only affects nuclei with $A>1$, 
  and so protons ($A=1$) are only affected by absorption
  through photo-pion interactions. 
These effective path lengths 
  are distinguished as dashed lines 
  in {Figure}~\ref{fig:H_1_z0_all}, 
  and are estimated by approximating these stochastic absorption process as a continuous loss process. 
This allows an equivalent cooling rate to be estimated as  
    ${b^{A \pi}}_{\! A} \approx -\epsilon_A{\Lambda^{A \pi}}_{\! A}$ 
    for pion-production, 
    or as ${b^{\rm sp}}_{\!A} \approx -\epsilon_A{\Lambda^{\rm sp}}_{\!A}$ f
    or photo-spallation\footnote{This is only suitable as an approximation, and is not strictly valid 
  as the fractional energy loss in a particular interaction 
  could be substantial \citep[see e.g.][]{Owen2018MNRAS}. 
Thus, particle energy losses are stochastic in nature, which introduces an additional statistical broadening to the absorption locations of UHE CRs impacted by this process. This broadening can change the mean path length by a factor of a few, and is particularly severe at high energies where the analytical path-length expression derived from the continuous approximation 
   begins to differ substantially 
   from that computed using a more appropriate numerical approach~ \citep{Dermer2009book}.}.  
Although approximated as continuous losses 
for the comparative purpose 
  of {Figures}~\ref{fig:H_1_z0_all} and~\ref{fig:z3z0compare}, our later calculations fully account for photo-pion and photo-spallation processes as absorption.


\section{Source populations} 
\label{sec:source_model} 

We adopt a parametrised model to describe the injection rate of CRs:
\begin{align}
\label{eq:astrophysical_injection}
  {\mathcal{Q}^{\rm a}}_{\! A}(z,\epsilon_A, A)
   & ={Q^{\rm a}}_{\! A}(1+z)^{-3} \nonumber \\ 
    & =\mathcal{C}_0(1+z)^{-3} \psi_{\rm x}(z)\;\!\frac{\mathrm{d}}
    {\mathrm{d}\epsilon_A}
    \left(\frac{\mathrm{d}{n}^{\star}}{\mathrm{d}t} \right) \ , 
\end{align} 
where ${\mathcal{Q}^{\rm a}}_{\!A}$ and ${Q^{\rm a}}_{\!A}$ 
  are the comoving and physical source terms, respectively. 
The normalisation $\mathcal{C}_0$ is discussed in 
 {Section}~\ref{subsec:cr_source_norm}. 
 $\psi_{\rm x}(z)$ 
  is the evolutionary function 
  describing the distribution of a CR source population 
  (the specific model is denoted by ${\rm x}$) 
  with respect to redshift $z$ and is discussed in {Section}~\ref{subsec:z_dists}, 
  and $\mathrm{d} n^{\star}/ \mathrm{d} t$ is the volumetric injection rate of CRs by a source in that population.

\subsection{Redshift distribution}
\label{subsec:z_dists}

Four source populations 
  are considered, and defined to a maximum redshift $z_{\rm max} = 3$. This maximum redshift is chosen because it would be extremely unlikely 
    that UHE CRs could reach us  
    from greater distances, given their suppression 
    by photo-hadronic interactions 
    with the CMB 
    and radiation emitted from astrophysical objects 
    during their propagation in intergalactic space. 
Each of the source population models we consider 
  follows a different redshift evolution. The first model (referred to as the star formation rate, or SFR, model) is based on the redshift evolution of cosmic star formation (also adopted by~\citealt{Muzio2019PRD}), and takes the form: 
\begin{equation}
\label{eq:z_SFR}
    \psi_{\rm SFR}(z) = \psi_{\rm SFR}^{0} \;\! \frac{(1+z)^{k_1}}{1+[(1+z)/k_2]^{k_3}} \ ,
\end{equation}
  where $k_1 = 2.7$, $k_2 = 2.9$ and $k_3 = 5.6$ 
  are empirical fit-parameters inherited from the best-fit function of~\cite{Madau2014ARAA}  
  \citep[see also][]{Robertson2015ApJ}\footnote{Alternative forms are adopted by some other researchers, for example, see~\cite{Wang2011ApJ, Batista2019JCAP} and~\cite{Palladino2020MNRAS}, 
 which used the redshift distribution 
 of \cite{Yuksel2008ApJ} instead.}.  

Our second model choice is a parameterised redshift distribution of gamma-ray bursts (and is referred to as the GRB model hereafter). It follows the form presented in \cite{Wang2011ApJ}: 
\begin{equation}
\label{eq:z_GRB}
    \psi_{\rm GRB}(z) = \psi_{\rm GRB}^{0} \;\! (1+z)^{k_4} \;\! \psi_{\rm SFR}(z) \ , 
\end{equation}
     where $k_4 = 1.4$. 
This is based 
   on indications from {\it Swift} observations  
  that the GRB rate 
  is enhanced relative to the SFR
  at high-redshift (\citealt{Le2007ApJ}; see also~\citealt{Yuksel2007PhRvD})\footnote{Note that alternative models are used in other studies to represent a GRB redshift distribution, for instance \cite{Batista2019JCAP} 
  instead adopt the GRB source redshift distribution 
  of~\cite{Wanderman2010MNRAS}.}. 

  We introduce a third model to describe an AGN source population (referred to as the AGN source model hereafter).
For this, we adopt the parametrisation of \cite{Hasinger2005A&A}   
\citep[see also ][]{Ahlers2009PhRvD,Wang2011ApJ,Batista2019JCAP}, 
\begin{equation}
\label{eq:z_AGN}
    \psi_{\rm AGN}(z) = \psi_{\rm AGN}^{0} 
\left\{
    \begin{array}{ll}
        (1+z)^{k_5} \hspace{1cm}  & (z < z_1)\  \\
        {z_2}^{k_5} \hspace{1cm} & (z_1 \leq z < z_2)\  \\
        {z_2}^{k_5}\cdot {z_2}^{z_2-z} \hspace{1cm} & (z\geq z_2)\   
    \end{array} 
\right. \ , 
\end{equation}  
  where $k_5 = 5.0$, $z_1=1.7$, $z_2=2.7$, which is 
 largely consistent 
  with the results in 
  the later survey studies on AGN populations, e.g.  
  \cite{Silverman2008ApJ,Ajello2012ApJ} 
  \citep[which were both adopted by][in the study of spectral and evolutionary properties of UHE neutrinos]{Jacobsen2015MNRAS},  
  and \cite{Ueda2014ApJ}. 
There are various uncertainties 
  in modelling other AGN contributions to the UHE CR flux.   
While it is unlikely 
  that the results obtained from our parametrised model 
  would need substantial revision,   
  these uncertainties, e.g., those concerning the AGN evolution, 
  could have detectable effects  
  on the spectral properties of UHE CRs at Earth.  
  
\begin{figure}
    \centering
    \includegraphics[width=\columnwidth]{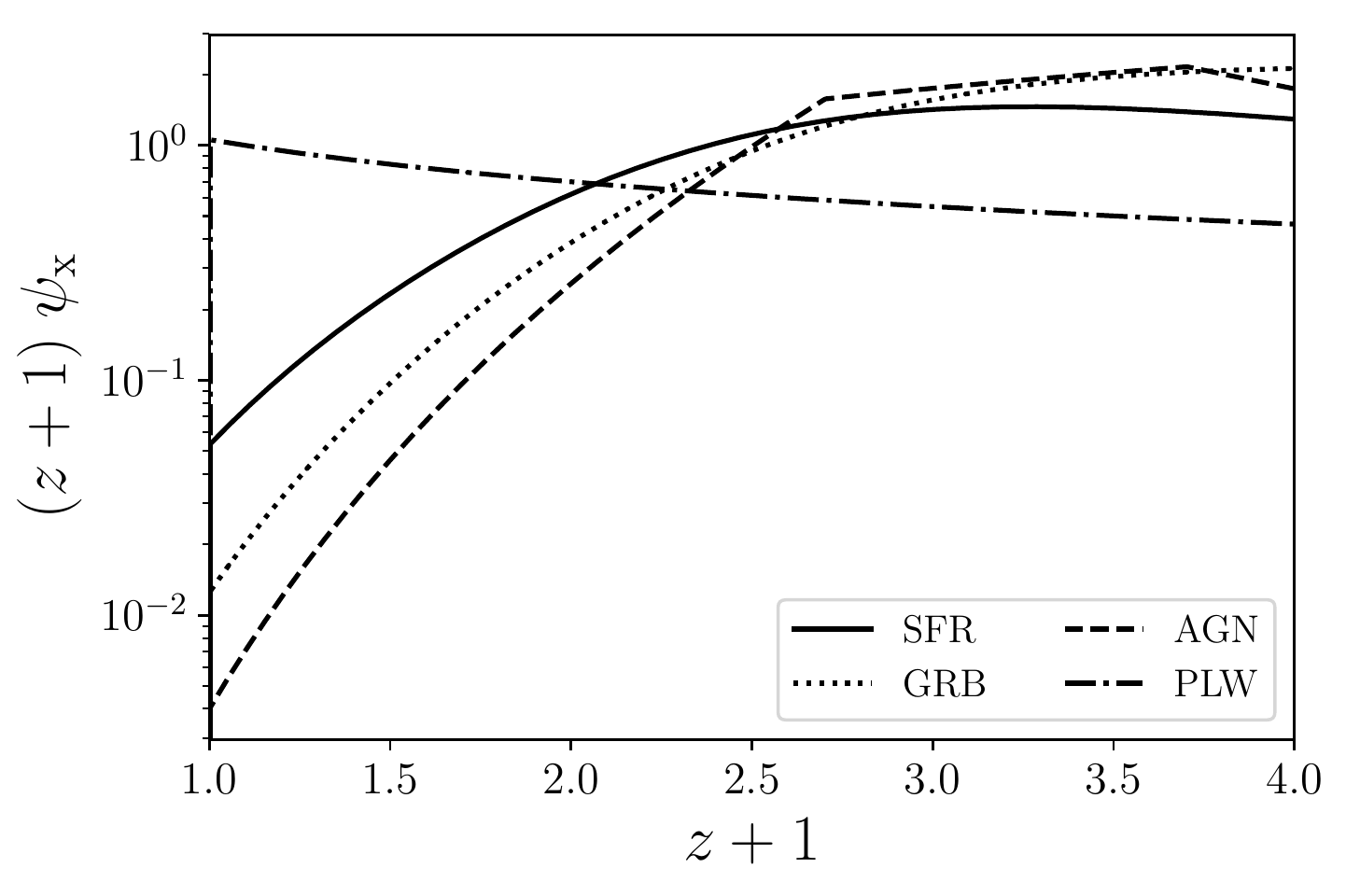}   
        \caption{Redshift distributions of 
        the four source models,  
        SFR, GRB, AGN and PLW.   
        Their corresponding normalized redshift evolution functions are given in {Equations} (\ref{eq:z_SFR}),  
        (\ref{eq:z_GRB}), (\ref{eq:z_AGN}), and (\ref{eq:z_PLW}) 
        respectively.}
  \label{fig:redshift_distributions}
\end{figure}

As the origins of UHE CRs are still to be determined, 
  some studies adopted a simple power-law parametrisation of the redshift evolution 
  of the source model, in the form $(1+z)^{k_{\rm PLW}}$, 
  where $k_{\rm PLW}$ is a fit parameter to be determined 
  \citep[e.g.][]{Taylor2015PhRvD, Aab2017JCAP, Jiang2020arXiv}. 
Although our other three source models are more physically-motivated, 
 for comparison we also consider 
  such a power-law parametrisation 
  (referred to as the PLW model), defined as   
\begin{equation}
\label{eq:z_PLW}
    \psi_{\rm PLW}(z) = \psi_{\rm PLW}^{0} 
    \;\! (1+z)^{k_{\rm PLW}}  \ , 
\end{equation} 
 with $k_{\rm PLW} = -1.6$, 
 the best-fit value obtained by~\cite{Batista2019JCAP}.  
In Equations (\ref{eq:z_SFR}) - (\ref{eq:z_PLW}), 
  $\psi_{\rm SFR}^{0}$, $\psi_{\rm GRB}^{0}$, 
   $\psi_{\rm AGN}^{0}$ and $\psi_{\rm PLW}^{0}$ 
   are the normalisation constants computed by integrating the respective source  
   distribution over redshift up to 
   a cutoff ($z_{\rm max} = 3$). 
 Their resulting values are presented  
   in {Table}~\ref{tab:parameters}, and 
 the redshift evolutionary histories for our four source models are shown in {Figure}~\ref{fig:redshift_distributions}.

\begin{deluxetable*}{lcccc}
\tablecaption{
Summary of parameter choices adopted for each of the four redshift source distribution models.}   
\label{tab:parameters}  
\tablecolumns{5}
\tablenum{1}
\tablewidth{0pt}  
\tablehead{
\colhead{Model} &
\colhead{ Normalization $\psi_{\rm x}^{0}$} &
\colhead{ Spectral index $\alpha$ } & 
\colhead{ ${\rm log}(R_{\rm max}/V)$} &
\colhead{ $\rho_{\rm CR}$/$10^{48}$ erg Mpc$^{-3}$ yr$^{-1}$} 
}
\startdata
    (1) SFR & $\psi_{\rm SFR}^{0}= 0.054$ & $-1.3$ & 18.2 & 0.5 \\
    (2) GRB & $\psi_{\rm GRB}^{0}= 0.013$ & $-1.5$ & 18.2 & 2.0 \\
    (3) AGN & $\psi_{\rm AGN}^{0}= 0.0041$ & $-1.0$ & 18.2 & 0.04 \\
    (4) PLW & $\psi_{\rm PLW}^{0}= 1.1$ & $+1.0$ & 18.7 & 15.0 \\
\enddata
\tablecomments{See text for the choices of the model parameters.} 
\end{deluxetable*}
  
  \begin{figure*}
    \centering
    \includegraphics[width=0.95\textwidth]{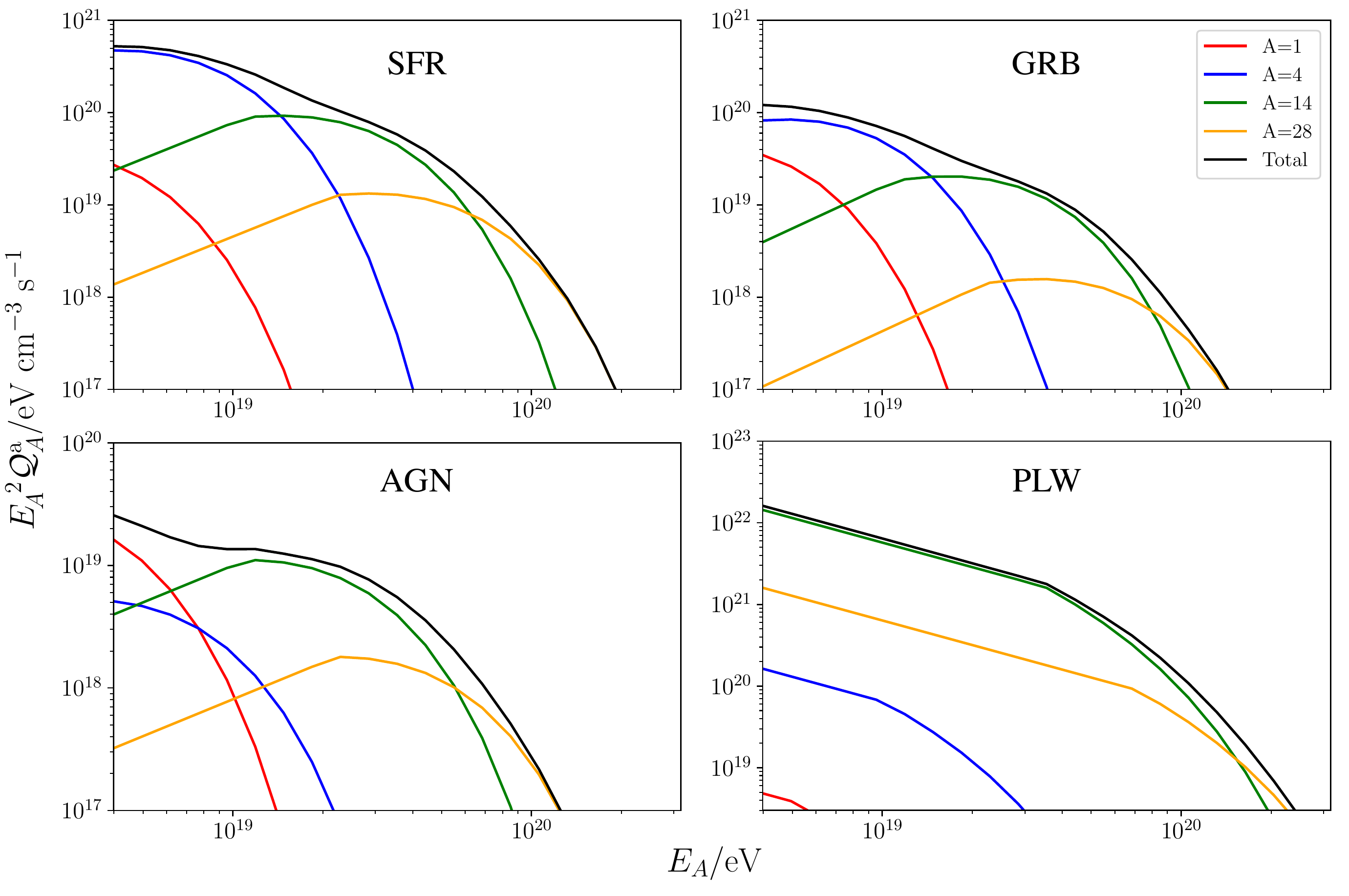}   
    \caption{Injection spectra at $z=0$ 
        for the four UHE CR source models,   
        SFR, GRB, AGN and PLW  
        as described in {Section} \ref{sec:source_model}.  
        ${\mathcal{Q}^{\rm a}}_{\!A}$ 
         is the comoving source term   
         that specifies the injection of CRs from 
         the astrophysical sources, 
         as sepcified in 
         Equation (\ref{eq:astrophysical_injection}).  
         The injection composed of four species,  
         with A=1 ($^1$H), 4 ($^4$He), 14 ($^{14}$N)
         and 28 ($^{28}$Si), are shown. 
         Their abundances are listed in 
         {Table}~\ref{tab:abundances} 
         (which are the best fit values from 
         \citealt{Batista2019JCAP}). }
  \label{fig:combined_sources}
\end{figure*}

\subsection{Cosmic ray luminosity density}
\label{subsec:cr_source_norm}

The normalisation term $\mathcal{C}_0$ 
 in Equation (\ref{eq:astrophysical_injection})  
  can be determined from 
   the UHE CR luminosity density,
  $\rho_{\rm CR}$.
This is the rate of UHE CR energy density generated by the source population and, 
  at $z=0$, it may be expressed as 
\begin{align}
\label{eq:overall_normalisation}
  \rho_{\rm CR} 
    &= \mathcal{C}_0\sum_{A}
   \int_{\epsilon_{\mathrm{min}}}^{\epsilon_{\mathrm{max}}} 
 \mathrm{d}\;\!\epsilon_A \;\! 
  \epsilon_A\;\! \psi_{\rm x}(z)\;\!  
  \frac{\mathrm{d}}{\mathrm{d}\;\!\epsilon_A}
  \left(\frac{\mathrm{d}\;\!{n}^{\star}}{\mathrm{d}\;\!t} \right)
  \biggr \vert_{z=0}  
\end{align}
\citep[cf.][]{Jiang2020arXiv}.  
 The exact value of $\rho_{\rm CR}$ is poorly constrained, however it may be roughly estimated by dividing the measured UHE CR energy density at $z=0$,    
  around $10^{54}~{\rm erg}~{\rm Mpc}^{-3}~{\rm dex}^{-1}$
  \citep[see][]{Aab2020PhRvL},  
  by a typical CR energy loss time,  
  $\ell/c \sim 1~{\rm Gpc}/c \sim 1~{\rm Gyr}$ (cf.~{Figure}~\ref{fig:H_1_z0_all}), giving $\sim 10^{45}\;\! \text{erg} \;\! \text{Mpc}^{-3} \;\! \text{yr}^{-1}$. This is comparable with more rigorous calculations \citep[e.g.][]{Berezinsky2006PRD,Aloisio2007APh,Wang2011ApJ,Jiang2020arXiv}, although the exact number is dependent on assumed model parameters -- in particular, the lower energy cut-off of the CR spectrum,    
   and the redshift distribution of the sources 
   \citep{Wang2011ApJ,Jiang2020arXiv}. 
This variation is evident, 
  as shown in the studies of \cite{Berezinsky2006PRD} 
  and \cite{Aloisio2007APh}, 
  where substantially larger values than our estimate, 
  of $3.5\times 10^{46}~\text{erg} \;\! \text{Mpc}^{-3} \;\! \text{yr}^{-1}$ 
  and $7\times 10^{46}~\text{erg} \;\! \text{Mpc}^{-3} \;\! \text{yr}^{-1}$ 
  (at energies above $10^{18}~{\rm eV}$) respectively, are derived     
  by assuming that there is 
  no redshift dependence in the UHE CR source populations.  
By contrast, much lower values, 
   in the range $0.2-2\times 10^{44}~\text{erg} \;\! \text{Mpc}^{-3} \;\! \text{yr}^{-1}$
   (at $3.16\times 10^{19}~{\rm eV}$)  
   were obtained by \cite{Jiang2020arXiv},  
   who adopted a negative redshift evolution for the UHE CR sources population. 
In \cite{Wang2011ApJ}, 
   a range of values 
   depending on the assumed source model and energy cut-off are presented: 
for energies above $10^{18}~{\rm eV}$, 
   they found the source luminosity density 
   to be $\sim 0.3-1\times 10^{45}~\text{erg} \;\! \text{Mpc}^{-3} \;\! \text{yr}^{-1}$,  
   for a source redshift distribution following that in  
   \citealt{Yuksel2008ApJ} 
   (corresponding to the SFR scenario in this work); $\sim 6\times 10^{44}~\text{erg} \;\! \text{Mpc}^{-3} \;\! \text{yr}^{-1}$ 
   assuming a redshift distribution following the AGN scenario    
   as in this work -- cf Equation (\ref{eq:z_AGN}); 
   and $\sim 0.8-3\times 10^{45}\;\!\text{erg}\;\!\text{Mpc}^{-3} \;\! \text{yr}^{-1}$ 
   assuming a source distribution 
   following that of~\cite{Yuksel2007PhRvD} 
   (cf. the GRB source scenario considered here).  
  While we do not perform detailed model fitting, we set the UHE CR luminosity density for each of the source models to give a reasonable flux at $z=0$ when compared to Pierre Auger Observatory data~\citep{Valerio2019APoS} -- see {Figure}~\ref{fig:combined_spectra_out} and 
  {Figure}~\ref{fig:PAO_compare}. The adopted values 
  (at energies above $5\times 10^{18}\;\!\text{eV}$) 
  are shown in  {Table}~\ref{tab:parameters}
  for each of the source models, 
  and these were used to set the normalisation $\mathcal{C}_0$  
  in Equation~(\ref{eq:overall_normalisation}) above.

\subsection{Source spectrum}
\label{subsec:source_spectrum}
  
We model the energy spectrum of the injected CRs 
  using the spectral form:    
\begin{align}%
\label{eq:normalisation_injection}
\frac{\mathrm{d}\;\!n^{\star}}{\mathrm{d}\;\!\epsilon_A} 
 & \propto f_{A} \left(\frac{\epsilon_A}{\epsilon_{\rm min}}\right)^{-\alpha} 
\left\{ 
\begin{array}{ll}
1 & ( x<1)  \\
\exp \left(1-x\right) & (x\geq 1) \
\end{array}\right.   \   
\end{align} 
\citep{Taylor2015PhRvD,Batista2019JCAP,Jiang2020arXiv}, 
where $x = {\epsilon_{\rm A} m_{\rm e} c^2}/{Z_A R_{\max }}$,
   $R_{\rm max}(V)$ is the rigidity and the energy limits    
  $\epsilon_{\mathrm{min}} m_{\rm e} c^2$ and 
   $\epsilon_{\mathrm{max}} m_{\rm e} c^2$ retain their earlier values. 
  Our parameter choices for the spectral index $\alpha$ and rigidity are shown in {Table}~\ref{tab:parameters}. 
Our adopted values 
  are comparable  
  to those derived from fits to Pierre Auger Observatory data  
  in the analysis of \cite{Batista2019JCAP}. 
  We note that $\alpha$ is strongly dependent on the adopted source population's redshift evolution 
  \citep[see][]{Taylor2015PhRvD}, 
  leading to substantial differences 
  in our choice of parameter value between the four models. 

   \begin{deluxetable*}{lccccc}
\tablecaption{
Abundance fractions adopted in the model 
for the four different 
UHE CR source redshift distributions.     
\label{tab:abundances} 
} 
\tablecolumns{6}
\tablenum{2}
\tablewidth{0pt}
\tablehead{ 
\colhead{ $\psi_{\rm x}(z)$ model} &
\colhead{ $f_{\rm 1}$ ($^1$H)  } &
\colhead{ $f_{\rm 4}$ ($^4$He)} &  
\colhead{ $f_{\rm 14}$ ($^{14}$N)} &
\colhead{ $f_{\rm 28}$ ($^{28}$Si)} & 
\colhead{ $f_{\rm 56}$ ($^{56}$Fe)}   
}
\startdata
    (1) SFR       &  0.1628 & 0.8046 & 0.0309 & 0.0018 & 0.0  \\
    (2) GRB       &  0.5876 & 0.3973 & 0.0147 & 0.0004 & 0.0  \\
    (3) AGN       &  0.8716 & 0.0778 & 0.0469 & 0.0038 & 0.0  \\
    (4) PLW &  0.0003 & 0.0101 & 0.8906 & 0.0990 & 0.0  \\
\enddata 
\end{deluxetable*}
   
\subsection{Injected composition}
\label{subsec:inj_comp}
   
We adopt a simplified CR composition source model, where the full range of injected species are represented by the abundances of $^1$H, $^4$He, $^{14}$N, $^{28}$Si and $^{56}$Fe. The individual component spectral forms are indicated in {Figure}~\ref{fig:combined_sources}, and their total contribution is normalised such that $f_{1}+f_{4}+f_{14}+f_{28}+f_{56}=1$.
The fitted composition fraction values obtained by 
\cite{Batista2019JCAP} 
  are adopted for species abundances 
  for each of the four source populations (SFR, GRB, AGN and PLW). The CR spectral models used in \cite{Batista2019JCAP}  
  and this work are slightly different,  
  but the fitted values given in \cite{Batista2019JCAP} 
  are still a reasonable choice for our model.
  Note that we calculate the production 
of all secondary nuclear CR species between $^{1}$H and $^{56}$Fe 
formed as the primary UHE CRs propagate and interact 
with cosmological radiation fields.
  The abundance fractions of the species used in our calculations 
  are shown in {Table}~\ref{tab:abundances}. 


\section{Results and Discussion} 
\label{sec:results}

\subsection{Cosmic ray spectrum \& composition}
\label{sec:results_spectrum}

\begin{figure*}
    \centering
    \includegraphics[width=0.95\textwidth]{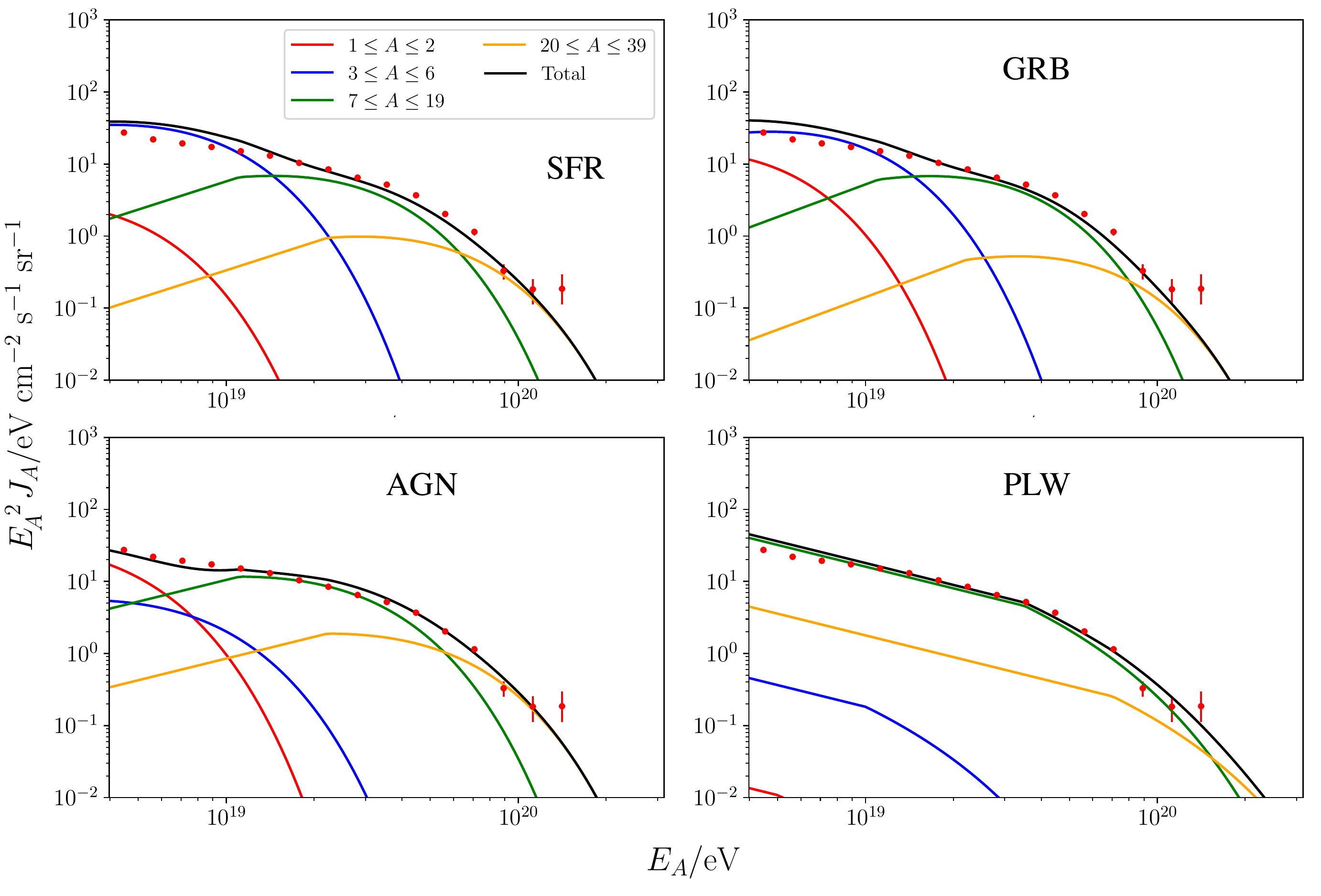}   
        \caption{Flux spectra of CRs 
        for the four source models,  SFR, GRB, AGN and PLW, 
        propagated to $z=0$.  
    The flux spectrum shown for the assigned range of $A$ 
      has accounted for the production of 
        secondary particles along the CR propagation.  
    The data (discrete red data points with error bars) 
       obtained by the Pierre Auger Observatory 
       \citep{Valerio2019APoS} 
       are shown for comparison.  
   The heaviest species 
     of nucleus injected by the sources considered here 
     is $^{28}$Si 
     in all the four models, i.e.  
     all the nuclei have mass number ${\rm A}\leqslant 28$. 
  The contribution from particles with $40 \leqslant A\leqslant 56$ 
    is therefore zero. 
} 
\label{fig:combined_spectra_out}
\end{figure*}

The CR spectrum, expected to be observed at $z=0$, 
  was computed by solving the transport equation  
  (Equation (\ref{eq:transport_equation_steadystate}) 
  in Section \ref{sec:cosmological_prop})
  numerically 
  using a Runge-Kutta integrator\footnote{An adaptive Runge-Kutta Fehlberg 4th order scheme~\citep{Press1992_book} was found to be sufficient.}.  
Cases for the four source populations, 
  SFR, GRB, AGN and PLW, 
  were calculated, with 
the numerical integration proceeding    
   from $z_{\rm max}=3$ to $z_{\rm min}=0$,  
   with an explicit treatment of 
   the injection of the CR compositions 
   and of the evolution of the source populations along $z$. 
The computed CR flux and composition at $z=0$ 
   took account of CR propagation and interaction effects, including 
   cooling, absorption and disintegration/spallation 
   (outlined in 
   {Section}~\ref{sec:propagation_interactions}). 
   The results are 
   are presented in {Figure}~\ref{fig:combined_spectra_out}. 
  
The production of secondary nuclei by photo-spallation 
  when CRs propagate in intergalactic space 
  drives the evolution of the composition of CRs, 
  with heavy nuclei gradually being eroded 
  to spawn lighter nuclear species, 
  i.e those of smaller atomic numbers.    
When the CRs are injected in our model, 
  they are comprised of five nuclear species 
  (i.e. $^1$H, $^4$He, $^{14}$N, $^{28}$Si and $^{56}$Fe) 
  as specified by the source models 
   (see {Section}~\ref{subsec:inj_comp}).
Other species are produced via spallation 
  as the CRs propagate from their sources to $z =0$. 
This evolution is captured   
  in {Figure}~\ref{fig:combined_spectra_out}, 
  which shows the compositions in four broad bands 
  to represent the binned distribution of nuclear species 
  when the CRs complete their journey to reach $z=0$.

When setting the abundance of $^{56}$Fe to be zero, 
  the heaviest species injected by the source model 
  (see {Section}~\ref{sec:source_model})  
  is $^{28}$Si ($A=28$), so no CR nuclei would have $A>28$.  
As the CRs propagate,  
  nuclei initially having $A =28$ 
  would become nuclei with $A\leq 28$.  
The four broad-band composition spectra in 
  {Figure}~\ref{fig:combined_spectra_out}  
  and the corresponding four injection species spectra 
  in {Figure}~\ref{fig:combined_sources}  
  have strikingly similar morphology and relative strengths. 
The flux in the band $1\leq A \leq 2$ is marginally higher, 
  relatively, than the injected H fraction by the source model, 
   while that in the and $3\leq A \leq 6$, $7\leq A \leq 19$ and $20\leq A \leq 39$ is marginally lower than the injected fractions 
   of $^{4}$He, $^{14}$N and $^{28}$Si respectively. 
At $z=0$ the flux in each band is dominated 
  by the injected source species,  
  indicating that 
  the evolution of species in an ensemble of CR nuclei 
  propagating over vast distances through intergalactic space 
  is only marginal 
  (subject to the effects of cosmological expansion, cooling, absorption, 
  and photo-spallation).  
We may therefore conclude that, in the absence of other additional factors,  
  CR species arriving at $z=0$ are overwhelmingly primary CRs 
  rather than secondaries re-processed by photo-spallation. 
This result is not particularly surprising  
  given that the injection process persists to $z=0$, 
  and the rate of photo-spallation interactions 
  is not comparable to the injection of primary CRs by the sources 
  for the four populations considered.  
It also explains 
  the limited evolution  
  seen between the injected source spectrum 
  and the post-propagation spectrum, 
  with cooling rates being easily overcome by new injections 
  at all energies 
  (see {Figure}~\ref{fig:H_1_z0_all}, which  
  also clarifies that 
  absorption processes dominate over any cooling process 
  for CRs with energy above $10^{20}~{\rm eV}$). 
  
\begin{figure}
    \centering
    \includegraphics[width=\columnwidth]{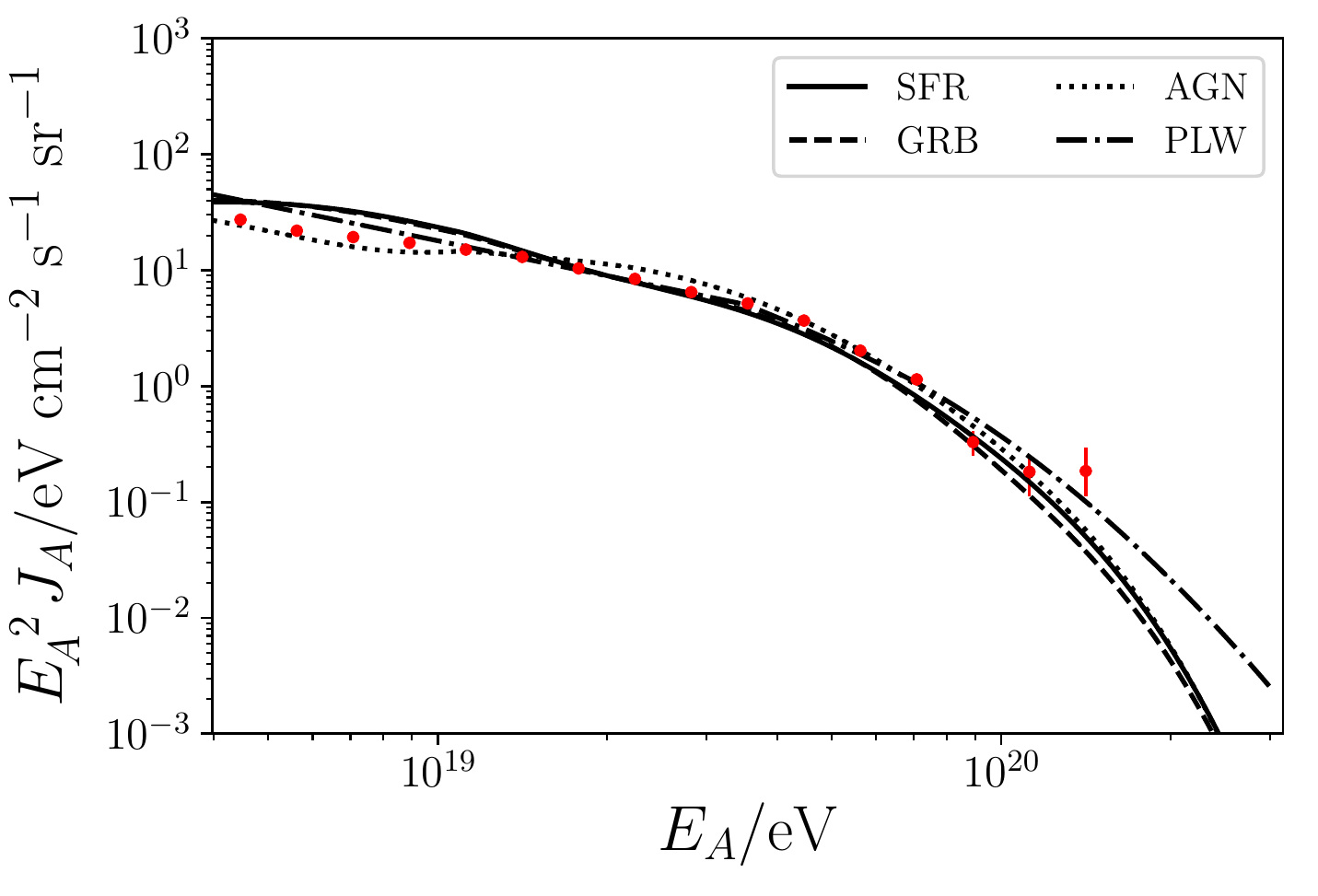}   
        \caption{Flux spectra of CRs for the source models,  
          SFR, GRB, AGN and PLW. 
    In each model,  
        the flux is the total value 
        obtained by summing the contributions of 
        all the species considered in the calculations.  
     The data (discrete red data points with error bars) 
       obtained by the Pierre Auger Observatory 
       \citep{Valerio2019APoS} 
       are shown for comparison. 
        }
  \label{fig:PAO_compare}
\end{figure}

Our calculations for all considered models 
  yield similar spectra at $z=0$ to that obtained by observations, 
  and this agreement is evident 
  (see {Figure}~\ref{fig:combined_spectra_out})  
  when comparing the computed total flux spectra 
  with data obtained by the Pierre Auger Observatory (PAO)   
  \citep{Valerio2019APoS}. {Figure}~\ref{fig:PAO_compare} 
  shows a comparison 
  between the total flux spectra obtained for 
  the four models, SFR, GRB, AGN and PLW.  
The AGN source population model 
  was found to produce (summed) spectra 
  best matching the observation,  
  while the flux levels  
  in all four model cases 
  are similar to the PAO observations 
  and other studies \citep[e.g.][]{Batista2019JCAP}. We note that an improved fit cannot be achieved with the parameter range considered in our models. Model refinement to obtain a better fit with available data is left to future studies.
  
\begin{figure}
    \centering
    \includegraphics[width=\columnwidth]{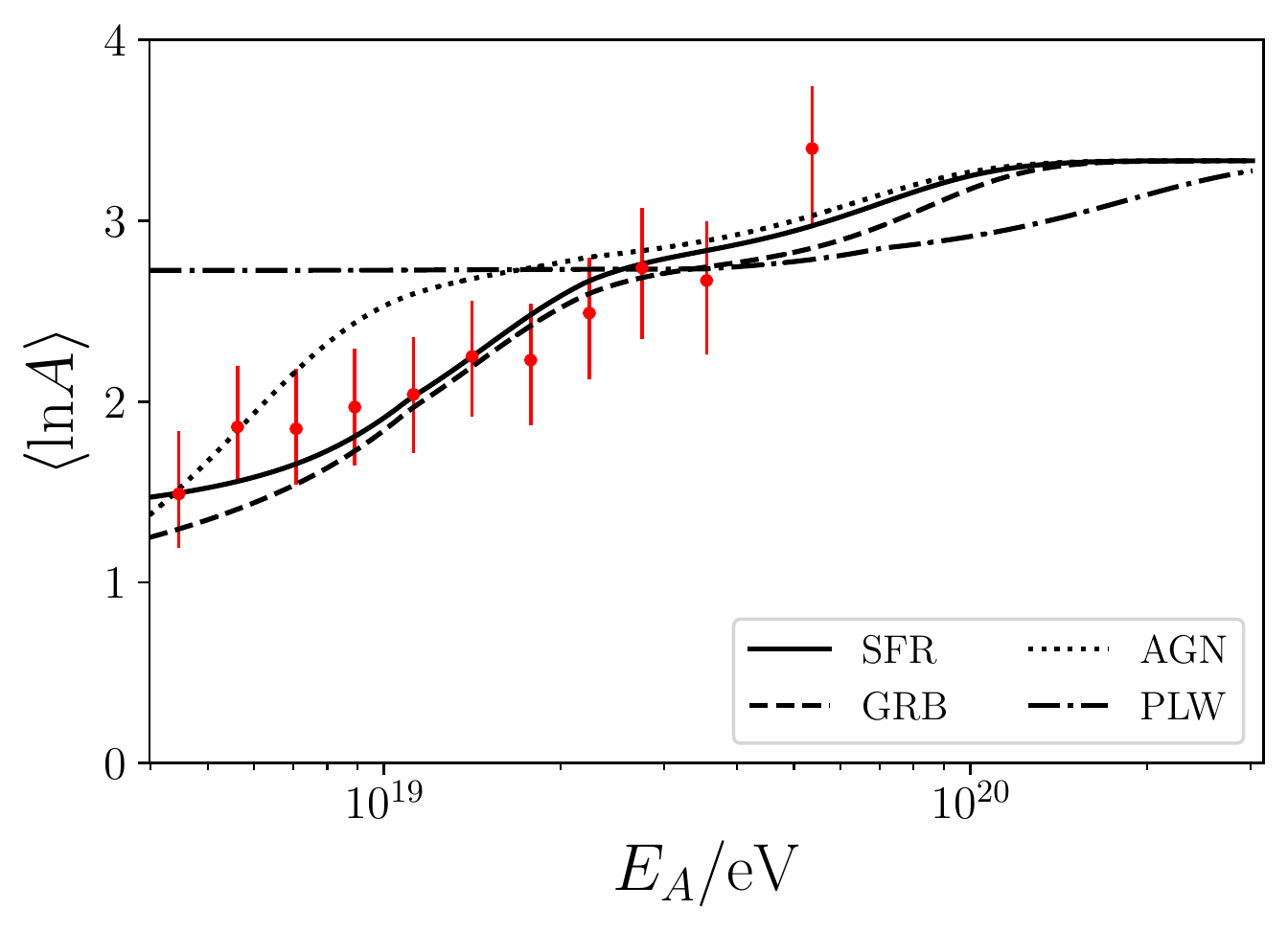}   
        \caption{Average UHE CR mass composition, $\langle {\rm ln} A \rangle$, for the four source classes, compared to data obtained with the Pierra Auger Observatory~\citep{PAO2013JCAP, Yushkov2019PoS} when adopting the Sibyll 2.3c hadronic interaction model~\citep{RiehnPoS2017}.}
  \label{fig:PAO_comp_compare}
\end{figure}

  Figure~\ref{fig:PAO_comp_compare} further compares
  the average UHE CR spectral mass composition, $\langle {\rm ln} A \rangle$, for the four source classes with data~\citep{PAO2013JCAP, Yushkov2019PoS}. 
  This shows a clearer distinction between the different source classes, and also clearly rules out 
  the PLW model as parametrised with Equation~(\ref{eq:z_PLW}). The AGN model is also disfavored in this comparison (particularly at around $E_{\rm A} \sim 10^{19}\;\!{\rm eV}$), however possible variations in the adopted hadronic model\footnote{The data shown are computed using the Sibyll 2.3c hadronic interaction model~\citep{RiehnPoS2017}, however other models are available~\citep[e.g.][]{Ostapchenko2011PhysRevD, Pierog2015PhysRevC}, and these yield slightly different $\langle {\rm ln} A \rangle$ spectral shapes and normalisations.}, or the possibility of combined source classes with a refined AGN-type component may offer scope for an improved fit.
 
\begin{figure*}
    \centering
    \includegraphics[width=0.95\textwidth]{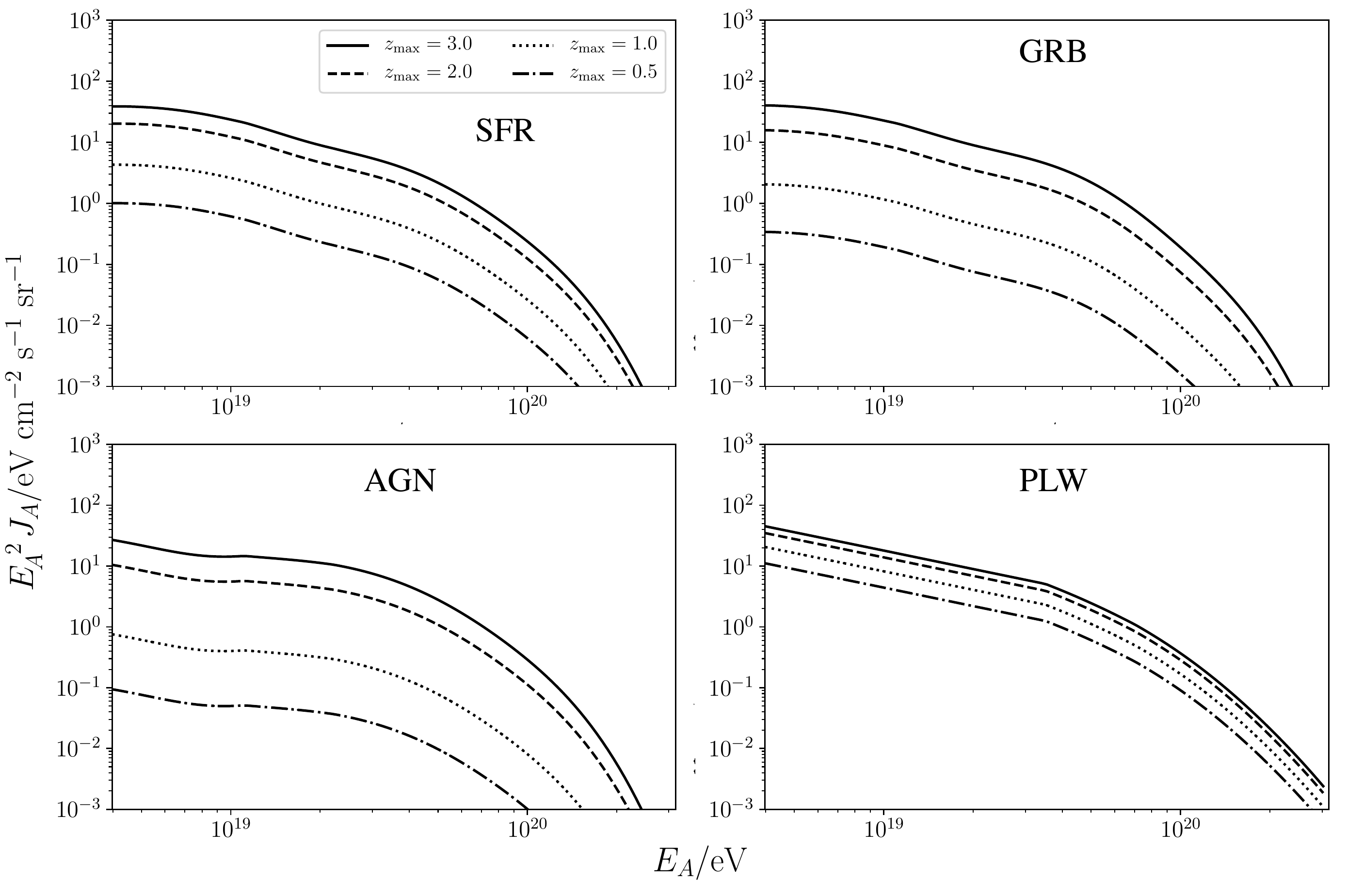}
        \caption{Total flux spectra of CRs propagated to $z=0$ 
         for the four source models.  
         In each model, 
          the cases for CR injection truncated 
            at redshifts $z_{\rm max}=0.5$, 1.0, 2.0 and 3.0 
          (curves from bottom to top) are shown. 
        Comparing the flux spectra for different redshift truncation     
          reveals that the contributions from distance sources 
          are significant.  
        } 
  \label{fig:zmax_spectra}
\end{figure*}

\subsection{Contributions from distant UHE CR sources} 
\label{sec:CR_anisotropy}

Anisotropy in the arrival directions 
  of UHE CRs above  $8\times 10^{18}~{\rm eV}$,   
  at a significance $>5.2 \sigma$ 
  was recently reported 
  \citep[see][]{Abreu2010APh, Aab2014ApJ, Aab2015ApJ, Aab2017Sci, Aab2018ApJ, Abbasi2018ApJ}. 
There are also indications of correlations 
  with certain candidate high-energy cosmic accelerators 
  \citep[see][]{Abraham2007Sci, Abreu2010APh}, 
  which include 
  AGN detected in $\gamma$-rays and X-rays 
  \citep{Nemmen2010ApJ, Terrano2012ApJ}\footnote{Note that this is in contrast with the finding of \cite{Mirabal2010MNRAS} \citep[see also][]{Alvarez2016JCAP}.}, 
  as well as source distributions 
  tracing large-scale structures, 
  with a source population being biased 
  with respect to the distribution of galaxies 
  \citep{Kashti2008JCAP}. 
A CR `hotspot' 
  was recently reported by \cite{Abbasi2014ApJ}.  
Studies by \cite{Fang2014ApJ} and \cite{He2016PhRvD} 
  showed that this hotspot is consistent 
  with a single source  
  rather than a chance signal.
Clusters of excess arrivals of UHE CRs above an isotropic background  
  from the direction of 
  Centaurus~A\footnote{Centaurus A has been considered as a source able to accelerate CRs to energies of $\sim 10^{20}\;\!{\rm eV}$ 
  at distances above 100 kpc from its core \citep{Peer2012JCAP}.} 
   have been found in some analyses~\citep{Fargion2011_proc, Kim2013JKPS, Kim2013ApJ}\footnote{Accounting also for a smearing effect due to intergalactic magnetic fields of order 10 nG~\citep{Kim2013ApJ}.}. 
 However, there is no clear detection of 
   excesses around other nearby radio galaxies,  
   e.g.~Centaurus B and Virgo A 
   \citep[see][]{Fraija2019JCAP,Kobzar2019MNRAS}. 

If UHE CRs are produced 
  by sources located nearby to our Galaxy with a fraction originating from within the conventional GZK horizon, as well as sources at large (cosmological) distances, then their arrival flux could have two components:  
  an isotropic component  
  corresponding to unresolved distant sources 
  and an anisotropic component 
  associated with relatively nearby sources. 
The reported anisotropies detected in UHE CRs 
  are consistent with this scenario,  
  and consideration of a foreground and background component in the UHE CR flux 
  was shown in \cite{Kim2011JCAP}   
  to improve anisotropy correlations
  with the positions of potential nearby sources.  
This is complimented 
  by the blind test performed by 
  \cite{Rubtsov2012JETPL} 
  for verifying 
  the hypothesis that AGN 
  were the origin of UHE CRs,  
  using event sets from Yakutsk, AGASA and HiRes.   
Their findings were consistent with a random background 
  \cite[see also][which also found 13 UHE CR events detected by HiRes in the northern hemisphere to be consistent with isotropy]{Abbasi2008APh}. 

\begin{figure*}
    \centering
    \includegraphics[width=0.75\textwidth]{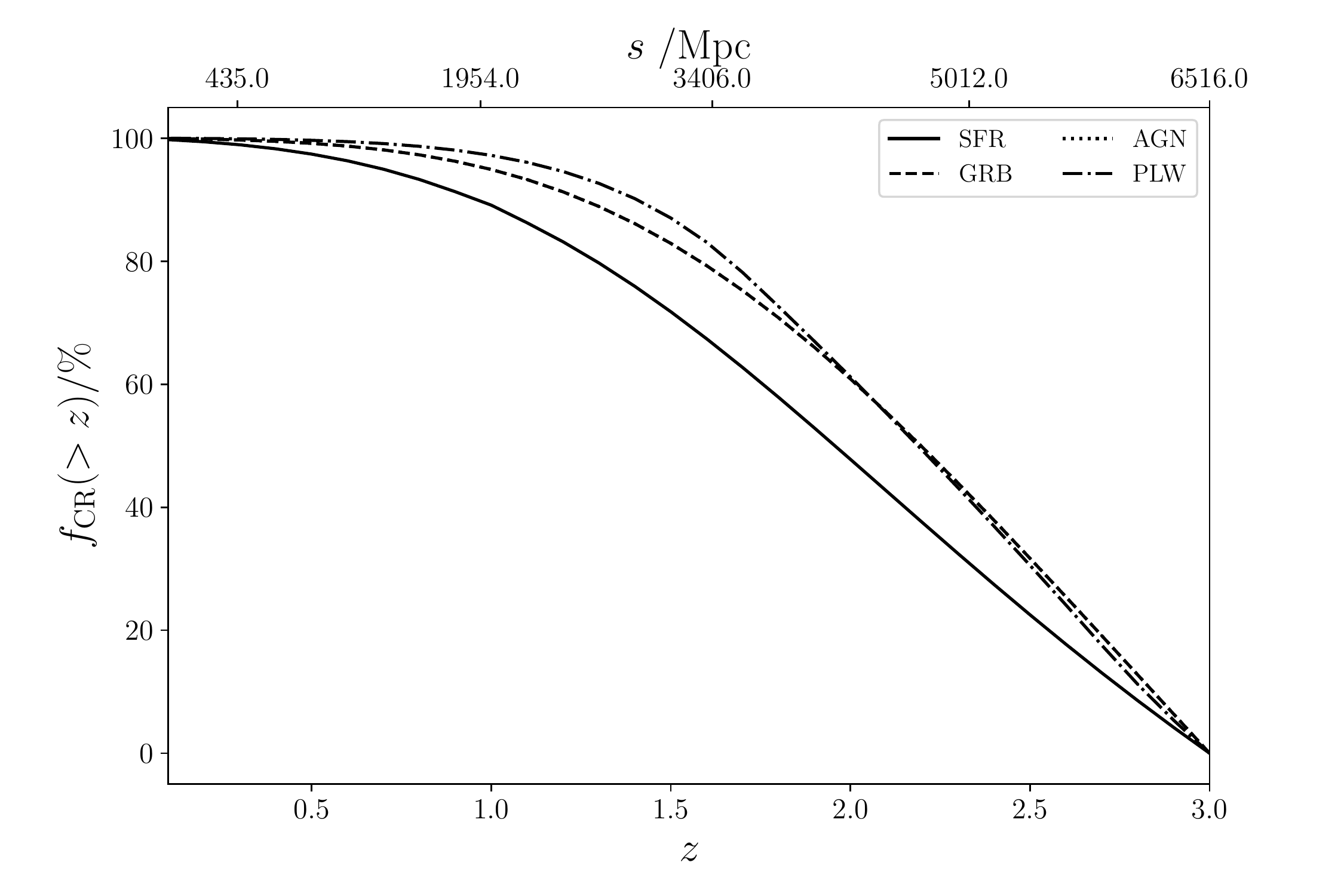}   
        \caption{The fraction of CRs 
          that originate from redshift higher than $z$,   
          $f_{\rm CR}(>z)$, for the four source models. 
        The bottom abscissa shows the redshift $z$, 
          and the top abscissa the corresponding comoving distances,  
          for the cosmological model with 
          $\Omega_{\rm m,0} = 0.315\pm0.007$, 
          $\Omega_{\rm r,0} \approx 0$,  
          $\Omega_{\rm \Lambda,0} = 0.685\pm0.007$, 
          and $h=0.673\pm0.006$ 
          (see {Section} \ref{sec:cosmological_prop}). 
         The curves for the GRB and AGN models  
           are almost indistinguishable,  
           a consequence of the redshift evolution of GRB and AGN,   
           in that most contributions from 
           their respective sources 
           are located at similar redshifts 
           (see {Figure} \ref{fig:redshift_distributions}). 
}
\label{fig:cr_distances}
\end{figure*}
 
Here, we assess the fraction of UHE CR flux that could originate at large distances, from sources located 
  beyond a GZK `horizon' distance of a few tens Mpc 
   \citep[see e.g.][]{Kachelries2019PrPNP}, 
   and determine their contribution to a UHE CR background. 
Distant CR sources 
  are homogeneously distributed across the celestial sphere
  over a range redshifts. 
Regardless the fraction of their flux which can survive to reach Earth, CRs originating from these sources 
  should not, at least in principle, 
  show resolvable anisotropy.   
To assess the relative contributions to the CRs 
  observed at $z=0$ 
  from the CRs initially produced by distant sources 
  at different redshifts, 
  we computed the flux spectra (for each of the source models)    
  with several assigned $z_{\rm max}$, 
  taking values from 3.0, 2.0, 1.0 and 0.5. 
  The spectra computed at $z=0$ for four source models are presented in {Figure}~\ref{fig:zmax_spectra} for four values of $z_{\rm max}$. This shows that changing either the source model or $z_{\rm max}$ leads to discernible changes in the spectra at $z=0$. 
The contribution of distant source populations is clearly substantial 
  and naturally accounts for the emergence 
  of an isotropic (background) component 
  in the observed UHE CR flux on Earth.

The significance of the distant sources' contribution  
  varies among the different source evolution models, 
  which can be seen from comparing 
  the set of spectra of different $z_{\rm max}$   
  shown in different panels 
  of {Figure} \ref{fig:zmax_spectra}. 
The amount of reduction 
  between the spectrum from sources 
  below $z=3$ to that from sources below $z=2$  
  accounts for a large fraction of the total CR flux 
  arriving at $z=0$, 
  which is at a level between 30$-$50\% 
  in the SFR, GRB and AGB models.
The PLW model however, has a relatively smaller reduction 
  in the flux for the same redshift interval,   
  implying that CR flux is dominated 
  by sources at lower redshifts in that case. 

{Figure}~\ref{fig:cr_distances} shows 
  the fractional contribution from sources  
  above a given redshift, $f_{\rm CR}(>z)$,
  for the four source models. 
Most noticeably, 
  the curves for the GRB and AGN models 
  are practically indistinguishable.   
The morphology of the $f_{\rm CR}(>z)$ 
  curve is a manifestation 
  of the redshift locations 
  where a specific population of sources 
  has the most significant contribution. 
The dominant redshift locations 
  of the GRB and AGN CR sources 
  are very similar to one another 
  (cf. the curves of these two source populations  
  in {Figure}~\ref{fig:redshift_distributions}, 
  particularly above $(z+1)\approx 2.5$).  
This, together with the  
  lacking of substantial numbers of GRB and AGN 
  below $(z+1) \sim 2.5$ 
  leads to their almost identical $f_{\rm CR}(>z)$ curves.   
Figure \ref{fig:cr_distances} 
  indicates that, for all the source population models considered,  
  a large faction of UHE CR flux observed at Earth 
  would originate 
  from regions 
  far beyond the conventional GZK horizon 
  of a few tens Mpc, and would not be attributed to local sources. 
If the UHE CRs are predominantly produced 
  as a consequence of star formation, 
  which is represented by the SFR model here,  
  then $\sim\;\!50$\% of the CRs 
  could originate from as far as $z\sim 2$ or above. 

These findings are robust.  
The results obtained from our calculations  
  are not sensitive 
  to the assumed composition of the injected UHE CR flux.   
The outcome is similar  
  even if the CR injections are restricted to a single species 
  (of any of the five nuclei considered).  
This is in line with the chemical evolution 
  due to photo-spallation processes 
  being insignificant, 
  otherwise, 
  the composition evolution of CRs 
  would be strongly manifested 
  in the CR energy spectra of the species observed at Earth.



\section{Summary and Conclusions}
\label{sec:conclusions} 

In this work we investigated the UHE CR flux and spectra at $z=0$ 
  with the explicit considerations of 
  (i) the distribution of candidate sources in the cosmological context  
  (modeled as four different redshift-dependent populations, 
  denoted as SFR, GRB, AGN and PLW as described in previous sections),  
  (ii) the composition of the injected CR nuclear species  
  (specified by the abundance of $^1$H, $^4$He, $^{28}$Si and $^{56}$Fe species),    
  and (iii) the relevant radiative and cooling processes, 
  and particle absorption interactions 
  (which include photo-spallation for the CR nuclei 
  when interacting with CMB and EBL radiation fields). 
We solved the particle transport equations numerically 
  and determined 
  the evolution of CR properties, 
  in particular  
  the CR flux and spectra 
  and the compositions of species for each model. 
Our calculations have shown that 
sources at redshifts as high as $z\sim (2-3)$ 
  can contribute substantially 
  to UHE CRs detected on Earth, with different source classes being distinguishable using the average UHE CR mass composition spectrum at $z=0$. Comparison with average CR mass data obtained from the Pierre Auger Observatory allowed a strong contribution from a PLW-type source population (as parametrised with Equation~\ref{eq:z_PLW}) to be ruled out. 

Regardless of population class, we find that most of the UHE CRs from these distant source populations  
  are primary particles,  
  despite the large cosmological distances  
  that they have traversed. 
These UHE CRs are diffuse and isotropic, 
  as they are of cosmological origin. 
They constitute the isotropic background 
  on which an anisotropic UHE CR component 
  associated with the nearby CR accelerators 
  is superimposed.

\section*{Acknowledgements}
ERO is supported by 
  the Center for Informatics and Computation in Astronomy (CICA) 
  at National Tsing Hua University (NTHU), 
  funded by the Ministry of Education of Taiwan (ROC). 
YXJY is supported by a NTHU International Student Scholarship 
  and by a grant from the Ministry of Science and Technology of Taiwan (ROC), 
  109-2628-M-007-005-RSP (PI Prof.\!\! Albert Kong). 
This work used high-performance computing facilities 
  operated by
   CICA at NTHU. 
This equipment was funded by the Ministry of Education of Taiwan and 
  the Ministry of Science and Technology of Taiwan.
The authors thank the anonymous referee for their comments, 
  which substantially improved the manuscript. 
This research has made use of NASA's Astrophysics Data Systems. 

\appendix

\section{Extra-galactic background light}
\label{sec:appendixa}

\begin{deluxetable}{lccc}
\tablecaption{
Summary of parameter choices for $z=0$ EBL model, approximating an upper-estimate for the EBL using blackbody components~\citep{Hauser2001AR}. Dimensionless weights $f_{i,0}$ have been estimated from relative normalisations.}   
\label{tab:ebl_params}  
\tablewidth{\columnwidth}
\tablecolumns{4}
\tablenum{3}
\tablehead{
\colhead{Component$^{a}$} & 
\colhead{T/K} &
\colhead{$\Theta_i$} &
\colhead{$f_{i,0}$}
}
\startdata
    Dust & 62 & $1.1\times10^{-8}$ & $3.5\times10^{-7}$ \\
    UV/O (1) & 400 & $6.7\times 10^{-8}$ & $9.0\times10^{-12}$ \\
    UV/O (2) & 1,000 & $1.7\times10^{-7}$ & $4.5\times 10^{-13}$ \\
    UV/O (3) & 3,000 & $5.1\times10^{-7}$ & $1.4\times 10^{-14}$ \\
    UV/O (4) & 5,500 & $9.3\times10^{-7}$ & $6.0\times 10^{-16}$ \\
    UV/O (5) & 12,000 & $2.0\times10^{-6}$ & $6.1\times 10^{-17}$
\enddata
\tablenotetext{a}{Note that the UV/O components are not physically-motivated. These are adopted to provide a sufficient approximation to the EBL spectrum in the energy range of interest while affording analytical tractability.} 
\end{deluxetable}

Extra-galactic background light is comprised of radiation emitted from astrophysical objects. Its energy is mainly concentrated in two spectral peaks -- one at optical wavelengths, being broadly associated with stellar emission from within populations of galaxies, while the other is at infra-red wavelengths, and is presumably dominated by dust-reprocessed astrophysical emission (also mainly originating from within galaxies).

The EBL is well-studied observationally in the local Universe~\citep[see][]{Cooray2016RSOS}. However, its redshift evolution is less well-constrained~\citep[for EBL constraints from observations over a range of redshifts, see e.g.][]{Franceschini2017AA}, and is determined by the evolution of its 
astrophysical source populations. While many approaches have been taken to model the cosmological evolution of the EBL in detail, including forward-evolutionary models~\citep[e.g.][]{Kneiske2010A&A, Finke2010ApJ}, 
backward-evolutionary models~\citep[e.g.][]{Dominguez2011MNRAS, Stecker2012ApJ}, and semi-analytical models~\citep[e.g.][]{Gilmore2012MNRAS, Inoue2013ApJ_atten}, these are each subject to certain inherent assumptions and uncertainties leading to substantial variations in predictions between models and modeling approaches. 
Although the EBL has been shown to affect the cosmological propagation of UHE CRs~\citep[e.g.][]{Allard2005A&A, Aloisio2013APh}, and the exact form of model adopted has been demonstrated to have non-negligible effects on the spectrum of UHE CRs~\citep[e.g.][]{Aloisio2013APh},
we do not find it to have substantial impacts on the results 
of this study, with any EBL model (if reasonable) yielding comparable results. 
We therefore adopt a simple analytic representation of the EBL, comprised of six superposed blackbody components, as listed in Table 3. Of these, one is attributed to dust (infrared) emission and the remaining five are a non-physical approximation to the UV/optical part of the EBL that presumably arises from stellar emission. Of these five, only the components `UV/O (1)' and `UV/O (2)' fall within the photon energy range relevant to our calculations.\footnote{Our choice of multiple un-physical blackbody components for the UV/optical region of the EBL is adopted for analytical tractability in this work -- however, more physically-representative modified blackbody approximations for the EBL have been adopted in other works~\citep[e.g.][]{Dermer2009book}.}
The components $i$ each take the form
\begin{equation}
    n_i(\epsilon_{\rm ph}, z)
  = \frac{8\pi\;\! f_{i}(z)}{\lambda_{\rm c}^3}
    \frac{\epsilon_{\rm ph}^2}{\exp (\epsilon_{\rm ph}/{\Theta}_i)-1} \ ,   \label{eq:general_form_ebl}
\end{equation}
where terms retain their earlier definitions.
The redshift-dependent dimensionless weights of the components are
\begin{equation}
    f_{i}(z) = f_{i, 0} \; \begin{cases}
    (1+z)^{3.1} \ , \hspace{0.9cm} z<z_{\rm cut}\\
    (1+z_{\rm cut})^{3.1} \ , \hspace{0.5cm} z\geq z_{\rm cut}
    \end{cases} \ ,
\end{equation}
up to $z=3$, where $z_{\rm cut} = 1.4$, which follows the `baseline' EBL redshift evolution model in \cite{Aloisio2013APh}. 
More detailed investigation of the impact of the redshift evolution 
  of the EBL on UHE CR propagation falls beyond the scope of this paper, 
  and is more appropriate for a dedicated future study.
Temperature, energy density, dimensionless temperature $\Theta_i$ and $z=0$ dimensionless weights $f_{i,0}$ for each EBL model component $i$ are listed in Table 3. These choices have been shown to give reasonable consistency with local observational constraints~\citep{Hauser2001AR}. 
Note that, for comparison, in the case of the CMB, $f=1$ and $\Theta(z) = 4.58\times 10^{-10} \;\! (1+z)$.


\bibliography{references}{}
\bibliographystyle{aasjournal}



\end{document}